\documentclass[prd,aps,reprint,floatfix,superscriptaddress,nofootinbib,preprintnumbers]{revtex4-1}

\usepackage[retainorgcmds]{IEEEtrantools}
\usepackage{relsize}
\usepackage{amsmath}  
\usepackage{amssymb}  
\usepackage{amsfonts}
\usepackage[usenames,dvipsnames]{color}
\usepackage[pdftex]{graphicx}
\usepackage{indentfirst}
\usepackage[pdftex]{hyperref}
\usepackage[toc,page]{appendix}
\usepackage[format=plain,justification=centerlast,labelformat = parens]{subcaption}
\usepackage[format=plain,justification=centerlast,labelformat = parens]{caption}

\newcommand{\be}{\begin{equation}}
\newcommand{\ee}{\end{equation}}
\newcommand{\ba}{\begin{eqnarray}}
\newcommand{\ea}{\end{eqnarray}}

\newcommand{\en}{\nonumber\\}

\newcommand{\de}{\delta}

\newcommand{\kk}{\mathbf{k}}
\newcommand{\xx}{\mathbf{x}}
\newcommand{\rr}{\mathbf{r}}
\newcommand{\vs}{\mathbf{s}}
\newcommand{\yy}{\mathbf{y}}

\newcommand{\qq}{\mathbf{q}}

\newcommand{\ksub}{\kappa_{\rm sub}}
\newcommand{\bksub}{\bar{\kappa}_{\rm sub}}

\newcommand{\bnsub}{\bar{n}_{\rm sub}}

\newcommand{\aap}{Astron. Astrophys.}

\newcommand{\aapr}{Astron. Astrophys. Rev.}

\newcommand{\mnras}{Mon. Not. R. Astron. Soc.}

\newcommand{\jcap}{Journal of Cosmology and Astroparticle Physics}
\newcommand{\apjl}{The Astrophysical Journal Letters}

\begin{document}

\title{On the power spectrum of dark matter substructure in strong gravitational lenses}

\author{Ana Diaz Rivero}
\affiliation{Harvard University, Department of Physics, \\Cambridge, Massachusetts 02138, USA}

\author{Francis-Yan Cyr-Racine}
\affiliation{Harvard University, Department of Physics, \\Cambridge, Massachusetts 02138, USA}

\author{Cora Dvorkin}
\affiliation{Harvard University, Department of Physics, \\Cambridge, Massachusetts 02138, USA}

\date{\today}

%%%%%%%%%%%%%%%%%%%%%%%%%%%%%%%%%%%%%%%%%%%%%%%%%%%%
\begin{abstract}
\noindent Studying the smallest self-bound dark matter structure in our Universe can yield important clues about the fundamental particle nature of dark matter. Galaxy-scale strong gravitational lensing provides a unique way to detect and characterize dark matter substructures at cosmological distances from the Milky Way. Within the cold dark matter (CDM) paradigm, the number of low-mass subhalos within lens galaxies is expected to be large, implying that their contribution to the lensing convergence field is approximately Gaussian and could thus be described by their power spectrum. We develop here a general formalism to compute from first principles the substructure convergence power spectrum for different populations of dark matter subhalos. As an example, we apply our framework to two distinct subhalo populations: a truncated Navarro-Frenk-White subhalo population motivated by standard CDM, and a truncated cored subhalo population motivated by self-interacting dark matter (SIDM). We study in detail how the subhalo abundance, mass function, internal density profile, and concentration affect the amplitude and shape of the substructure power spectrum. We determine that the power spectrum is mostly sensitive to a specific combination of the subhalo abundance and moments of the mass function, as well as to the average tidal truncation scale of the largest subhalos included in the analysis. Interestingly, we show that the asymptotic slope of the substructure power spectrum at large wave number reflects the internal density profile of the subhalos. In particular, the SIDM power spectrum exhibits a characteristic steepening at large wave number absent in the CDM power spectrum, opening the possibility of using this observable, if at all measurable, to discern between these two scenarios.

\end{abstract}

\maketitle

%%%%%%%%%%%%%%%%%%%%%%%%%%%%%%%%%%%%%%%%%%%%%%%%%%%%
\section{Introduction}\label{sec:introduction}
%%%%%%%%%%%%%%

In our Universe, structure formation based on the Cold Dark Matter (CDM) paradigm \cite{1981ApJ...250..423D,Blumenthal:1982mv,Blumenthal:1984bp,Davis:1985rj} has been extremely successful at explaining the large-scale distribution of matter across cosmic times. On subgalactic scales however, assessing whether CDM provides a good fit to observations is significantly more difficult. On the one hand, baryonic processes can play an important role on these scales \cite{Brooks_Zolotov,Brooks_2013,Arraki,Onorbe_2015,Wetzel_2016,2016MNRAS.457.1931S,2017MNRAS.468.2283C,2017MNRAS.467.4383S,2017arXiv170103792G}, thus significantly affecting the dark matter distribution inside galaxies and their satellites, and making it difficult to compute robust theoretical predictions that can be compared to observations. On the other hand, star formation becomes increasingly inefficient in low-mass CDM halos \cite{Fitts:2016usl,2017MNRAS.467.2019R}, rendering their detection and characterization within the Local Group quite challenging.

Further complicating the picture is the fact that key aspects of the particle nature of dark matter might have important consequences on these subgalactic scales. For instance, significant dark matter free streaming \cite{Bond:1983hb,Dalcanton:2000hn,Bode:2000gq,Boyanovsky:2008ab,Boyanovsky:2011aa} or possible interactions with relativistic species  \cite{1992ApJ...398...43C,Boehm:2001hm,Ackerman:2008gi,Feng:2009mn,Kaplan:2009de,Aarssen:2012fx,Cyr-Racine:2013ab,Chu:2014lja,Buen-Abad:2015ova,Chacko:2016kgg,Ko:2016uft} at early times can  substantially reduce the number of low-mass subhalos orbiting a typical galaxy \cite{Buckley:2014hja,Schewtschenko:2014fca,Vogelsberger_2015}. In addition, dark matter self-interaction \cite{Spergel_1999,Yoshida00,Dave01,Colin02} could modify the density profile of main and satellite halos \cite{Vogels_2012,Rocha_2012,Peter:2012jh,Zavala_2013,Kaplinghat:2013xca,Vogelsberger_2015,Kaplinghat:2015aga} away from the standard CDM prediction  \cite{Navarro_1996}. Other dark matter particle candidates such as ultralight axions \citep{Hu_2000,Hui_2016} might also lead to interesting phenomenology on small scales (see e.g.~Refs.~\cite{Du_2016,Mocz:2017wlg}). 

Disentangling the impact of dark matter physics on structure formation from that of baryons is key to probing the fundamental nature of dark matter. While it is never entirely possible to neglect the influence of baryonic structures on the evolution of the small-scale dark matter distribution (see, e.g.~Ref.~\cite{2017arXiv170103792G}), it can be minimized by focusing our attention on the lowest mass subhalos present in galaxies. As mentioned above, these small subhalos are largely devoid of stars, which makes them less susceptible to baryonic feedback effects, while their abundance and internal structure are quite sensitive to the particle nature of dark matter, making them an important laboratory to test the consistency of the CDM paradigm on small scales. 

These dark subhalos could potentially be probed within the Local Group using detailed observations of tidal streams \cite{2014ApJ...788..181N,2016ApJ...820...45C,2016PhRvL.116l1301B,2016MNRAS.463..102E,2017MNRAS.466..628B} or the motion of stars within the Milky Way disk \cite{2012ApJ...750L..41W,Feldmann:2013hqa}. Beyond our local neighborhood however, gravitational lensing is the only technique capable of probing low-mass subhalos at cosmological distances from the Milky Way. In particular, galaxy-scale strong lensing systems in which a massive foreground galaxy is multiply-imaging a background source (such as another galaxy or a quasar) constitute ideal environments to study the cosmological population of low-mass dark subhalos. For instance, the study of flux-ratio anomalies in strongly lensed quasars \cite{Mao:1997ek,Metcalf2001,Chiba:aa,Metcalf:ac,Dalal:aa,Keeton:2003ab,Metcalf:2010aa} has lead to a measurement of the typical abundance of mass substructures within lens galaxies \cite{Dalal:2002aa}, and has also been used to put constraints on the position and mass of potential individual subhalos within the lens galaxies \cite{Fadely_2012,Nierenberg_2014,Nierenberg_2017}. 

Individual mass substructure can also be detected by carefully examining the surface brightness variation of extended lensed arcs and rings. This direct ``gravitational imaging'' \cite{Koopmans_2005,Vegetti:2008aa} has lead to the statistically significant detection of a few mass substructures with masses above $\sim10^8 M_\odot$ \cite{Vegetti_2010_1,Vegetti_2010_2,Vegetti_2012}. A somewhat similar technique using spatially resolved spectroscopic observations of gravitational lenses  \cite{Moustakas_Metcalf_2002,Hezaveh_2013} has also lead to the direct detection of a $\sim 10^9 M_\odot$ subhalo \cite{Hezaveh_2016_2}. Taken together, these measurements can be used to put constraints on the subhalo mass function (see, e.g.~Refs.~\cite{Vegetti:2009aa,Vegetti2014,Li:2015xpc}).

The main limitation of direct subhalo detection efforts is that only the most massive substructures lying within or very close to lensed arcs can be detected with large statistical significance. While not directly detectable, smaller mass substructures or those lying further away from lensed images could still potentially lead to observable effects on the lensing signal, especially on the relative arrival time delay between lensed images \cite{Keeton:2009ab,Cyr-Racine_2015}, but also on extended arcs. For instance, Refs.~\cite{Birrer2017,Brewer2015,Daylan:2017kfh} have recently proposed statistical techniques to harness the constraining power from these marginal detections on the properties and abundance of dark matter subhalos within lens galaxies.

Within the CDM paradigm, the subhalo mass function is expected to rise rapidly toward smaller masses \cite{Springel_2008}, implying that typical lensed images could be perturbed by a fairly large number of unresolved low-mass substructures. In this limit, it becomes somewhat impractical to phrase the perturbations to lensed images in terms of individual subhalos. A more fruitful approach in this case is to describe the substructure convergence field in terms of its $n$-point correlation functions. For CDM, the large number of small-mass subhalos contributing to the total substructure convergence field implies that the statistics of the latter should be nearly Gaussian. In this case, we expect the two-point correlation function (or its Fourier transform, the power spectrum) to dominate the statistical description of the substructure field. This last point was put forth in Ref.~\cite{Hezaveh_2014} to motivate an exploratory study of the detectability of the substructure convergence power spectrum within lens galaxies using the  Atacama Large Millimeter/submillimeter Array (ALMA). In practice, given that strong lensing is probing the matter density field deep in the nonlinear regime, we do not expect the substructure density field to be entirely Gaussian. Nevertheless, measuring the substructure power spectrum might still lead to important insights about the abundance and internal structure of subhalos within lens galaxies. 

Interestingly, Ref.~\cite{Hezaveh_2014} showed that it is possible, in principle, to measure the substructure convergence power spectrum by looking at the correlations of lensed image residuals, once a model image obtained from a purely smooth lens potential is subtracted from the  data.  They further showed that deep observations of strong gravitational lenses with ALMA could lead to 3-$\sigma$ detection of the nonvanishing amplitude of the substructure power spectrum (at least if there is abundant substructure, which is the case in CDM). Given that  such measurements might be possible in the near future, the immediate question that comes to mind is: \emph{What will we learn about low-mass subhalos from measuring the substructure convergence power spectrum?}

In this paper, we present some much-needed answers to this question. Using the standard halo model \cite{Cooray:2002dia} as our framework, we first develop a general formalism to compute the power spectrum of the convergence field on the lens plane due to substructure. We extend the initial approach presented in Ref.~\cite{Hezaveh_2014} to include subhalo populations that are not necessarily isotropic and homogeneous, and also take into account the 2-subhalo term. This formalism is developed in a way that makes it easy to change the statistical properties of the population as well as the intrinsic properties of subhalos, in order to facilitate its application to different dark matter scenarios. As an example we apply it to two different subhalo populations: one in which subhalos are modeled as truncated Navarro-Frenk-White (NFW) halos as would occur in standard CDM, and another one in which they are modeled as truncated cored halos as would happen in the presence of self-interacting dark matter (SIDM). We choose the latter because of there is evidence of cored density profiles in at least some of the Local Group satellites (e.g.~Refs.~\cite{Gilmore_2007, Walker_Penarrubia_2011, Amorisco_2013}). We then use these two examples as a springboard to discuss how the internal structure, statistical properties, and abundance of low-mass subhalos affect the shape and amplitude of the substructure convergence power spectrum.

This paper is organized as follows. In Sec.~\ref{sec:subs_stats} we present our halo model-based formalism to compute the substructure convergence power spectrum from first principles. In Sec.~\ref{sec:tnfw} we apply this formalism to study the 1- and 2-subhalo contributions to the substructure power spectrum from a population of truncated NFW subhalos. In Sec.~\ref{sec:sidm} we turn our attention to the substructure power spectrum in the presence of a population of truncated cored subhalos, highlighting along the way the differences from the NFW case. We finally discuss our findings and conclude in Sec.~\ref{sec:conclusions}. 

%%%%%%%%%%%%%%%%%%%%%%%%%%%%%%%%%%%%%%%%%%%%%%%%%%%%
\section{Substructure statistics within the halo model}\label{sec:subs_stats}

We work within the framework of the halo model \cite{Cooray:2002dia}, where all the dark matter is bound in roughly spherical halos. Within this model, the dark matter content of a typical lens galaxy is comprised of a smooth dark matter halo containing most the galaxy's mass, as well as a certain number of subhalos orbiting within the smooth halo. In the following, we will be concerned with these subhalos. 

%%%%%%%%%%%%%%%%%%%%%%%%%%%%%%%%%%%%%%%%%%%%%%%%%%%%
\subsection{Preliminaries: Subhalo statistics}

We work in projected two-dimensional (2D) space, with $\rr$ denoting the projected 2D vector in the plane of the sky. The total convergence at a given point $\rr$ on the lens plane is 
\be
\kappa_{\rm tot}(\rr)= \kappa_0(\rr) + \kappa_{\rm sub}(\rr),
\ee
where $\kappa_0$ denotes the contribution from the smooth lens model (dark matter + baryons) and $\kappa_{\rm sub}$ denotes that from the subhalos. Note that the convergence is nothing more than the projected mass along the line of sight $\Sigma$ in units of the critical density for lensing, $\kappa  \equiv \Sigma/\Sigma_{\rm crit}$, where $\Sigma_{\rm crit}$ depends on the angular diameter distance between the observer and the source $D_{\rm os}$, the observer and the lens $D_{\rm ol}$ and the lens and the source $D_{\rm ls}$:
\begin{align}
\Sigma_{\rm crit} = \frac{c^2 D_{\rm os}}{4\pi G D_{\rm ol} D_{\rm ls}}.
\end{align}
Here, $G$ is the gravitational constant and $c$ the speed of light. 

The convergence is also related to the projected Newtonian gravitational potential $\phi$ via the Poisson equation: $\triangledown^2\phi = 2 \kappa$.  According to the standard CDM model, a typical lens galaxy will contain a large population of subhalos, all of which contribute to $\kappa_{\rm sub}$ as:
\be\label{eq:ksub}
\kappa_{\rm sub}(\rr) = \sum_{i =1}^{N_{\rm sub}} \kappa_i (\rr-\rr_i, m_i, \qq_i),
\ee
where $\kappa_i$ and $\rr_i$ are the convergence and the position of the $i$th subhalo, respectively, $m_i$ is the total mass of the $i$th subhalo, and the $\qq_i$'s are sets of parameters that determine the internal properties of the $i$th subhalo. $N_{\rm sub}$ is the total number of subhalos contributing to the lensing convergence at position $\rr$. Note that in Eq.~\eqref{eq:ksub} we have taken advantage of the fact that the overall contribution of the subhalo population is equivalent to the sum of the effect of each subhalo, which follows from the linearity of Poisson's equation. Since the convergence profile of a subhalo is always directly proportional to the subhalo mass $m_i$, it is useful to define $\hat{\kappa}_i \equiv \Sigma_{\rm crit} \kappa_i/m_i$. The advantage of this notation is that $\hat{\kappa}_i$ obeys a very simple normalization condition
\be\label{eq:normalization_hat_kappa}
\int d^2\rr_i\,\hat{\kappa}_i(\rr_i,\qq_i)= 1,
\ee
independent of the value of $\qq_i$. Here, the integral runs over the whole lens plane. 

In general, it is impossible to know the mass, position, and internal properties of every subhalo within a lens galaxy. Instead, we would like to determine the ``ensemble-averaged" properties of gravitational lensing  observables given the statistical properties of subhalos, such as their mass function and spatial distribution. We shall denote by $\langle X \rangle$ the ensemble average of quantity $X$ over all possible realizations of the subhalo density field within a lens galaxy. On the other hand, the notation $\bar{X}$ will be used to denote the ``spatial" average of $X$ over a given area of the lens plane.

Let us assume that all the statistical properties of subhalos within a lens galaxy are captured  by a probability distribution function $\mathcal{P}(\rr,m,\qq)$. It is, in general, a very good approximation (see Refs.~\cite{Springel_2008,Fiacconi:2016cih}) to assume that the mass and projected position of a subhalo are uncorrelated. This allows us to write the overall distribution as a product of a mass and position probability distributions as follows:
\be\label{eq:product_probdist}
\mathcal{P}(\rr,m,\qq) = \mathcal{P}_{\rm r}(\rr)\mathcal{P}_{\rm m}(m)\mathcal{P}_{\rm q}(\qq|m,\rr)  ,
\ee
where we have taken into account that the intrinsic properties of a given subhalo likely depend on its mass and position within the lens galaxy. The distribution $\mathcal{P}_{\rm r}(\rr)$ contains all the information about the projected spatial distribution of subhalos within the host galaxy. Given a projected number density $n_{\rm sub}(\rr)$ of subhalos, the probability of finding a subhalo within an area $d^2\rr$ centered at position $\rr$ is
\be\label{eq:p_of_r}
\mathcal{P}_{\rm r}(\rr)d^2\rr = \frac{n_{\rm sub}(\rr)d^2\rr}{\int_{A} d^2\rr\, n_{\rm sub}(\rr) },
\ee
where $A$ is the area of the lens plane where we have sensitivity to substructures (see below). The denominator in Eq.~\eqref{eq:p_of_r} is just the total number of subhalos within the area $A$
\be\label{eq:normalization_nsub}
\int_{A} d^2\rr\, n_{\rm sub}(\rr) = N_{\rm sub} \equiv A\, \bar{n}_{\rm sub},
\ee
where $\bar{n}_{\rm sub}$ is the average number density of subhalos averaged over the whole area $A$. It is useful to write the subhalo number density as 
\be\label{eq:number_density_sub}
n_{\rm sub}(\rr) = \bar{n}_{\rm sub}\left(1 + \de(\rr)\right),
\ee
where $\de(\rr)$ is a stochastic random variable with $\langle \de(\rr)\rangle = 0$. Here, the $\de(\rr)$ field describes the fractional excess probability (compared to $ \bar{n}_{\rm sub}$) of finding a subhalo at position $\rr$. While any choice of $\de(\rr)$ fully specifies the probability density function $\mathcal{P}_{\rm r}(\rr)$ as per Eq.~\eqref{eq:p_of_r} statistically independent, we will, in general, be interested in ensemble-averaging over realizations of the $\de(\rr)$ field. 

Numerical studies \cite{Springel_2008,Fiacconi:2016cih} indicate that the 3D spatial distribution of subhalos near the central part of the host has a rather weak radial dependence. Taking into account projection effects and the fact that galaxy-scale strong lensing is mostly probing a small region near the projected center of the host, it is usually an excellent approximation to take $\langle n_{\rm sub}(\rr)\rangle = \bar{n}_{\rm sub}=$ constant.  

The subhalo mass probability distribution can be written as
\be\label{eq:dndM}
\mathcal{P}_{\rm m}(m)\equiv \frac{1}{N_{\rm sub}}\frac{dN_{\rm sub}}{dm}, 
\ee
where $d N_{\rm sub}/dm$ is the standard subhalo mass function. While our results are easily generalizable to any choice of mass function, we restrict ourselves to a power law mass function, $\mathcal{P}_{\rm m}\propto m^\beta$, for $m_{\rm low} < m < m_{\rm high}$. In the following, we assume that $\mathcal{P}(\rr,m,\qq)$ is normalized such that
\be\label{eq:normalization}
\int dm \, d^2\rr\,  d\qq \, \mathcal{P}(\rr,m,\qq) = 1,
\ee
which is trivially satisfied by Eqs.~\eqref{eq:p_of_r} and \eqref{eq:dndM}.

As in most lensing calculations in the literature, the calculations presented in the remainder of this paper assume that each subhalo represents an \emph{independent} draw from the $\mathcal{P}(\rr,m,\qq)$ probability distribution. We emphasize though that this does \emph{not} mean that we neglect spatial correlations between subhalos; these are fully encoded in our choice of $\mathcal{P}_{\rm r}(\rr)$. In this case, the probability distribution describing the properties of the whole subhalo population $\mathcal{P}_{\rm pop}$ can be factored out as a product of the probability distribution for single subhalos
\be
\mathcal{P}_{\rm pop} = \prod_{i  = 1}^{N_{\rm sub}} \mathcal{P}(\rr_i, m_i, \qq_i). 
\ee
We now have all the ingredients to perform ensemble averages over all possible realizations of a subhalo population. 

%%%%%%%%%%%%%%%%%%%%%%%%%%%%%%%%%%%%%%%%%%%%%%%%%%%%
\subsection{Ensemble-averaged substructure convergence}

It is instructive to first compute the mean ensemble-averaged substructure convergence on the lens plane $\bksub$. It is given by
\begin{align}
\bksub &= \frac{1}{A} \int d^2\vs\, \langle\ksub(\vs)\rangle \\
&=\frac{N_{\rm sub}}{A}\int dm_i\, d\qq_i \, \mathcal{P}_{\rm m}(m_i)\mathcal{P}_{\rm q}(\qq_i)  \en
&\qquad\qquad \times \int  d^2\vs \,d^2\rr_i  \, \kappa_i (\vs - \rr_i,m_i, \qq_i)\mathcal{P}_{\rm r}(\rr_i),\nonumber
\end{align}
where we used the fact that every term in the sum in Eq.~\eqref{eq:ksub} contributes equally to $\bksub$. The result is not surprising since it just states that the average convergence for the whole population of (statistically independent) subhalos is just $N_{\rm sub}$ times the average convergence of a single subhalo. Next, we note that the $\rr_i$ integral above is nothing more than the convolution of the subhalo density profile $\kappa_i$ with the spatial distribution $\mathcal{P}_{\rm r}$. Using the general result for the integral of a convolution,
\be
\int d^2\vs\, (f*g)(\vs) = \int d^2\vs \, f(\vs) \int d^2\rr \,g(\rr),
\ee
we obtain,
\begin{align}\label{eq:pract_to_phys}
\bksub &= \frac{N_{\rm sub}}{A\Sigma_{\rm crit}}\int dm_i \, \mathcal{P}_{\rm m}(m_i) m_i  \en
& = \frac{N_{\rm sub}\langle m\rangle}{A\Sigma_{\rm crit}},
\end{align}
where we used Eq.~\eqref{eq:normalization_hat_kappa}. In the above, we have introduced the notation 
\be
\langle m \rangle \equiv \int  dm_i \mathcal{P}_{\rm m}(m_i) m_i
\ee
to denote the average subhalo mass. We note that Eq.~\eqref{eq:pract_to_phys} is useful to relate $N_{\rm sub}$ and $A$ to the physically relevant quantities $\langle m\rangle$ and $\bar{\kappa}_{\rm sub}$. 

%%%%%%%%%%%%%%%%%%%%%%%%%%%%%%%%%%%%%%%%%%%%%%%%%%%%
\subsection{The power spectrum of the convergence field}

We now turn our attention to the computation of the two-point correlation function of the substructure density field, or its Fourier transform, the substructure power spectrum.  We emphasize that we do not assume here that the substructure convergence field is necessarily Gaussian. As such, we do not expect the power spectrum to characterize the substructure density field completely, and expect higher-point correlation functions to also contain nontrivial information. Nevertheless, the rapidly rising subhalo mass function toward the low-mass end in CDM models ensures that Gaussianity is a good first approximation \cite{Cyr-Racine_2015}. Importantly, the main contributors of non-Gaussianities to the substructure field are the most massive subhalos within the lens galaxy \cite{Hezaveh_2014}. Since we expect them to be directly detectable \cite{Vegetti_2012,Vegetti_2010_2, Hezaveh_2013,Hezaveh_2016_2}, we can limit their influence on the statistics of the $\kappa_{\rm sub}$ field by absorbing the most massive subhalos within the macrolens mass model $\kappa_0$.

To obtain a general expression for the substructure power spectrum $P_{\rm sub}(k)$, we first compute the lens plane-averaged connected two-point correlation function $\xi_{\rm sub}(\rr)$ of the substructure convergence field $\kappa_{\rm sub}$. To simplify the derivation and avoid clutter, we first focus exclusively on performing the spatial averages encoded in the probability distribution $\mathcal{P}_{\rm r}(\rr)$. The averages over the subhalo mass and internal properties will be restored at the end of the calculation. The substructure convergence two-point function takes the form
\begin{align}
\xi_{\rm sub}(\rr) &\equiv \frac{1}{A}\int d^2\vs\, \int \prod_i d^2\rr_i  \mathcal{P}_{\rm r}(\rr_i) \\
&\qquad \times (\ksub(\vs)-\bksub)( \ksub(\vs+\rr)-\bksub).\nonumber
\end{align}
Substituting Eq.~\eqref{eq:ksub} in the above and using the normalization condition given in Eq.~\eqref{eq:normalization}, we obtain
\begin{align}\label{eq:xi_sub_first}
A \xi_{\rm sub}(\rr) & = \sum_i  \int d^2\vs \,d^2\rr_i \kappa_i(\vs - \rr_i) \kappa_i(\vs +\rr - \rr_i)\mathcal{P}_{\rm r}(\rr_i)\en
&\qquad+\sum_i\sum_{j\neq i} \int d^2\vs \,d^2\rr_i  \,d^2\rr_j \mathcal{P}_{\rm r}(\rr_i) \mathcal{P}_{\rm r}(\rr_j)\en
&\qquad\qquad\qquad\times \kappa_i(\vs - \rr_i)\kappa_j(\vs+\rr - \rr_j)\en
&\qquad- \bksub \sum_i  \int d^2\vs \,d^2\rr_i \kappa_i(\vs - \rr_i) \mathcal{P}_{\rm r}(\rr_i) \en
&\qquad- \bksub \sum_i  \int d^2\vs \,d^2\rr_i \kappa_i(\vs +\rr - \rr_i) \mathcal{P}_{\rm r}(\rr_i)\en
&\qquad+ \bksub^2\int d^2\vs. 
\end{align}
The first term arises from ensemble-averaging over the spatial distribution of a single subhalo (the ``1-subhalo" term), the second term arises from averaging over pairs of distinct subhalos (the ``2-subhalo" term), while the last three terms ensure that we are computing only the connected part of the two-point function. In the language of the halo model, the 1-subhalo term refers to particles or mass elements within a same subhalo, while the 2-subhalo term is due to those in distinct subhalos. The 1-subhalo term is nothing else than the convolution of the subhalo density profile with itself
\begin{align}
& \int d^2\vs \,d^2\rr_i \kappa_i(\vs - \rr_i) \kappa_i(\vs +\rr - \rr_i)\mathcal{P}_{\rm r}(\rr_i)\en
&\qquad= \int d^2\xx \, \kappa_i(\xx) \kappa_i(\xx+\rr ) \en
& \qquad = (\kappa_i * \kappa_i)(\rr).
\end{align}
The 2-subhalo contribution contains $N_{\rm sub}(N_{\rm sub} - 1)$ identical terms which have the following form \cite{Sheth:2002cs}
\begin{align}
&\int d^2\vs \,d^2\rr_i  \,d^2\rr_j \mathcal{P}_{\rm r}(\rr_i) \mathcal{P}_{\rm r}(\rr_j)\kappa_i(\vs - \rr_i)\kappa_j(\vs+\rr - \rr_j)\en
&\qquad=\int d^2\xx \,d^2\yy \,\kappa_i(\xx)\kappa_j(\yy)  (\mathcal{P}_{\rm r}* \mathcal{P}_{\rm r})(\yy-\xx -\rr).
\end{align}
Using Eqs.~\eqref{eq:p_of_r} and \eqref{eq:number_density_sub}, the convolution of the subhalo's spatial distribution is
\begin{align}\label{eq:P_r_to_xiss}
(\mathcal{P}_{\rm r}* \mathcal{P}_{\rm r})(\rr) &=  \int d^2\vs\, \mathcal{P}_{\rm r}(\vs)\mathcal{P}_{\rm r}(\vs+\rr)\en
&=\frac{\bnsub^2}{N_{\rm sub}^2}\int d^2\vs\, (1+\de(\vs))(1+\de(\vs+\rr))\en
&=\frac{\bnsub}{N_{\rm sub}^2} \left( N_{\rm sub} + \bnsub\int d^2\vs\, (\vs)\de(\vs+\rr)\right)\en
&=\frac{\bnsub}{N_{\rm sub}} \left(1 + \xi_{\rm ss}(\rr)\right),
\end{align}
where we have identified the two-point subhalo correlation function $\xi_{\rm ss}(\rr)$, which encodes spatial correlation between pairs of distinct subhalos.  Finally, the three last terms of Eq.~\eqref{eq:xi_sub_first} all have the same form and lead to a net contribution of $-\bksub^2 A$. The connected two-point correlation function of the substructure convergence field thus takes the form
\begin{align}\label{eq:kappa_correlationfunction}
\xi_{\rm sub}(\rr) &= \frac{N_{\rm sub}}{A} (\kappa_i*\kappa_i)(\rr)\\
& + \frac{\bnsub N_{\rm sub}(N_{\rm sub}-1)}{AN_{\rm sub}}\int d^2\xx \,d^2\yy \,\kappa_i(\xx)\kappa_j(\yy) \en
&\hspace{3.5cm} \times \left(1 + \xi_{\rm ss}(\yy-\xx-\rr)\right) \en
& - \bksub^2. \nonumber
\end{align}
Noting that some of the integrals not involving $\xi_{\rm ss}$ in the second term exactly cancel the third term, we are left with
\begin{align}\label{eq:xi_sub_final}
\xi_{\rm sub}(\rr) &= \bnsub (\kappa_i*\kappa_i)(\rr)\\
& + \bnsub^2\int d^2\xx \,d^2\yy \,\kappa_i(\xx)\kappa_j(\yy)  \xi_{\rm ss}(\yy-\xx-\rr) \en
& -  \frac{\bnsub^2}{N_{\rm sub}}\int d^2\xx \,d^2\yy \,\kappa_i(\xx)\kappa_j(\yy)  \left(1 + \xi_{\rm ss}(\yy-\xx-\rr)\right).\nonumber
\end{align}
The first two terms correspond to the 1-subhalo and 2-subhalo terms, respectively, while the last term, suppressed by an extra factor of $N_{\rm sub}$, corresponds to the shot noise term, which only becomes important if the number of subhalos within the area of interest in the lens plane is small.

It is now straightforward to compute the convergence power spectrum by Fourier transforming Eq.~\eqref{eq:xi_sub_final}. Using the following Fourier transform conventions:
\begin{align}
\tilde{\kappa}(\kk) &= \int d^2\rr\, e^{-i\kk \cdot \rr} \hat{\kappa}(\rr), \\
\hat{\kappa}(\rr) &= \int \frac{d^2\kk}{(2\pi)^2}\, e^{i\kk \cdot \rr } \tilde{\kappa}(\kk)  ,
\end{align}
the convergence power spectrum takes the form
\begin{align}\label{eq:psub}
P_{\rm sub}(\kk) & = \int  d^2\rr\, e^{-i\kk \cdot \rr} \xi_{\rm sub}(\rr) \en
& = \bnsub |\tilde{\kappa}_i(\kk)|^2 \en
& \qquad +\bnsub^2 (1-\frac{1}{N_{\rm sub}})\tilde{\kappa}_i(\kk) \tilde{\kappa}_j^*(\kk)P_{\rm ss}(\kk),
\end{align}
where $\kk$ is the wavevector, and where we have used the convolution theorem to perform the Fourier transform. We note that the $\rr$-independent part of the last term in Eq.~\eqref{eq:xi_sub_final} contributes an unobservable zero-mode, which we dropped in the above. Here, $P_{\rm ss}(\kk)$ is the Fourier transform of the subhalo two-point correlation function $\xi_{\rm ss}(\rr)$. In the remainder of the paper we neglect the $1/N_{\rm sub}$ term in Eq. \eqref{eq:psub}.

Up to this point, the only assumptions underpinning our calculation of the substructure convergence power spectrum are the statistical independence of each subhalo within a lens galaxy, and the fact that the subhalo internal properties $\qq_i$ do not depend on the subhalo position $\rr_i$. We now introduce two simplifying assumptions:
\begin{itemize}
\item We take the subhalo convergence profile to be circularly symmetric, implying that $\tilde{\kappa}_i(\kk) = \tilde{\kappa}_i(k)$.
\item We assume that the subhalo two-point correlation function $\xi_{\rm ss}$ is homogeneous and isotropic, hence leading to $P_{\rm ss}(\kk) = P_{\rm ss}(k)$. 
\end{itemize}
Here, $k\equiv|\kk|$. While subhalos are generally triaxial, projection effects and ensemble-averaging over all possible orientations and sizes of the subhalos' ellipticity imply that the average convergence profile is close to circularly symmetric, hence our first assumption. Our second point amounts to assuming that the small area of the lens plane probed by strong lensing images is typical of other nearby lines of sight. With these assumptions, the Fourier transform of the subhalo convergence profile is
\begin{align}
\tilde{\kappa}(k) &= \int d^2\rr\, e^{-i\kk \cdot \rr} \hat{\kappa}(r)\en
& = 2\pi  \int dr\, r\, J_0( k \,r)  \hat{\kappa}(r),
\end{align}
where $J_0(x)$ is the 0th order Bessel function. 

The last step of the calculation is to reinstate the averages over subhalo mass and internal properties. We can write the total substructure convergence power spectrum as the sum of the 1-subhalo and 2-subhalo terms,
\begin{align}
P_{\rm sub}(k) &= P_{\rm1sh}(k) + P_{\rm2sh}(k),
\end{align}
where the 1-subhalo term $P_{\rm 1sh}(k)$ takes the form
\begin{align}\label{eq:1sh_constdist}
P_{\rm 1sh}(k) &= \frac{(2\pi)^2 \bksub}{\langle m\rangle \Sigma_{\rm crit}} \int dm \, d\qq \, m^2 \; \mathcal{P}_{\rm m}(m) \; \mathcal{P}_{\rm q}(\qq|m)\en
&\qquad \qquad \qquad \times \left[ \int dr \, r J_0(k\,r) \hat{\kappa}(r,\qq)\right]^2 
\end{align}
(the subscript $i$ has been dropped since it is now superfluous) and the 2-subhalo term takes the form
\begin{align}\label{eq:p2sh}
P_{\rm 2sh}(k) &= \frac{(2\pi)^2\bksub^2}{\langle m \rangle^2} P_{\rm ss}(k)\Bigg[\int dm \, d\qq \,m \, \mathcal{P}_{\rm m}(m) \, \mathcal{P}_{\rm q}(\qq|m)\en
&\qquad\qquad\qquad  \times   \int dr \, r J_0(k\,r) \hat{\kappa}(r,\qq) \Bigg]^2.
\end{align} 

The amplitude of the 1-subhalo term is approximately given by  $P_{\rm1sh}(k)\propto \bksub m_{\rm eff}$, where the quantity $m_{\rm eff}\equiv \langle m^2\rangle/\langle m\rangle$ has been referred to as the ``effective mass'' in the lensing literature \cite{1991A&A...250...62R,2003A&A...404...83N,Keeton_2009}. This specific mass scale constitutes the primary dependence of the substructure power spectrum on the subhalo mass function, so we expect it to be one of the most constrained quantities with actual observations. The amplitude of the 1-subhalo term can be approximated as $P_{\rm1sh}(k)\approx \bksub m_{\rm eff}/\Sigma_{\rm crit}$. For a typical gravitational lens with $0.003<\bksub<0.03$  \cite{Dalal:2002aa},  $m_{\rm eff}\sim 10^7 M_\odot$, and $\Sigma_{\rm crit}\sim 3 \times 10^9 M_\odot/{\rm kpc}^2$ (given our choices for the source and lens redshift), we thus expect
\be
P_{\rm1sh}(k)\sim 10^{-5}-10^{-4} \,{\rm kpc}^2
\ee
for scales larger than the typical size of a subhalo. On the other hand, the amplitude of the 2-subhalo term is approximately $P_{\rm 2sh}(k)\propto \bksub^2 P_{\rm ss}(k)$, with very little dependence on the subhalo mass function. Given that typically $\bksub \ll 1$ and that $P_{\rm ss}(k)$ can be important only on scales larger than the typical subhalo spatial separation, this term is generally subdominant compared to the 1-subhalo term, except maybe on larger scales, depending on the size of $P_{\rm ss}(k)$.

Having derived the general expression for the lens plane-averaged substructure power spectrum, we can now apply it to realistic subhalo populations by specifying the probability distributions $\mathcal{P}(\rr,m,\qq)$ and the subhalo convergence profile $\kappa(\rr,m,\qq)$.  For definiteness, we make the following choices throughout the rest of this paper whenever we present numerical results: we assume a lens galaxy at redshift $z=0.5$ with virial mass and radius $M_{\rm vir}=1.8 \times 10^{12}$ $M_{\odot}$, $R_{\rm max} = 409$ kpc, and Einstein radius $b=6.3$ kpc. We take the source to be at $z = 1$.

%%%%%%%%%%%%%%%%%%%%%%%%%%%%%%%%%%%%%%%%%%%%%%%%%%%%
\section{Truncated Navarro-Frenk-White subhalo population}
\label{sec:tnfw}

\subsection{Characteristics of the subhalo population}

In this section we compute the substructure power spectrum for a realistic population of smoothly truncated Navarro-Frenk-White subhalos. We are particularly interested in the strong lensing region, namely the region bounded more or less by the Einstein radius of the lens. Reference ~\cite{Cyr-Racine_2015} performed a detailed analysis of the statistics of subhalo populations in strong lenses by looking at both the ``local" (close to the Einstein radius of the host) and ``distributed" (extending past the host virial radius) populations of subhalos and looking at their relative effects on lensing observables such as the lensing potential, deflection, shear and convergence. They found that the substructure contribution at a typical image position is largely dominated by the local subhalos. 

\begin{figure}[t!]
\includegraphics[width=0.49\textwidth]{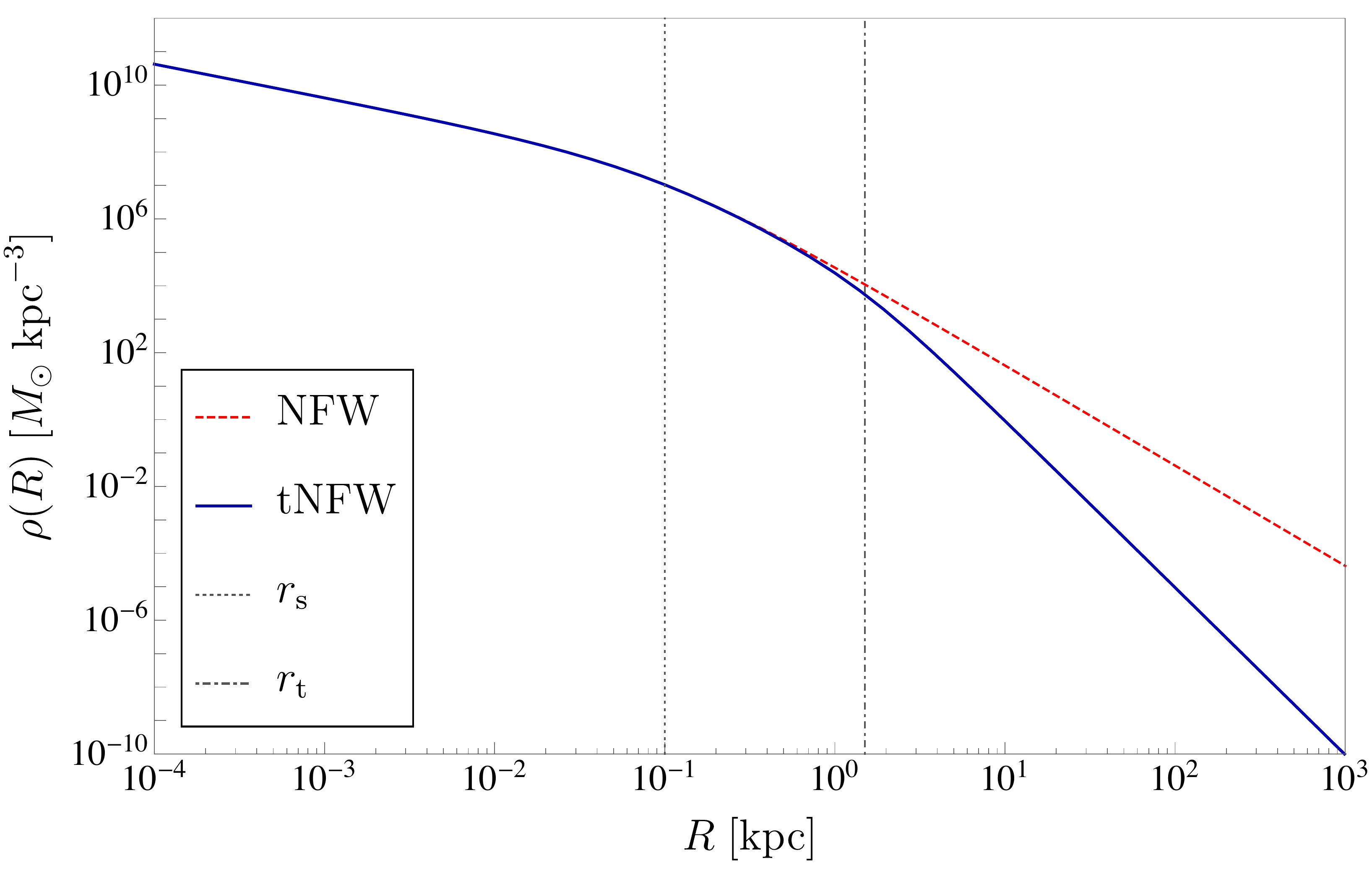}
\caption{Density profile for a regular NFW profile (dashed red) and a truncated NFW profile (blue) for $\tau = 15$ and $m = 10^6$ $M_{\odot}$. The dotted and dashed-dotted gray lines represent the scale and tidal radius, respectively.}\label{fig:cdm_density_profile}
\end{figure}

The NFW density profile \cite{Navarro_1996} has been found to provide a good fit to simulated CDM halos and is widely used to model the distribution of dark matter within galaxies and their satellites. This density profile (see Fig.~\ref{fig:cdm_density_profile}) has an inner slope that goes as $R^{-1}$ until it reaches the scale radius $r_{\rm s}$, where the slope steepens to $R^{-3}$. Formally, the NFW density profile leads to a divergent total subhalo mass. However, we expect tidal interactions to provide a finite truncation radius for a realistic subhalo orbiting within its host galaxy, hence leading to a finite subhalo mass. Here, we adopt the following truncated NFW profile (tNFW) \cite{Baltz_2007} for our subhalos:
\be\label{eq:3DNFW}
\rho_{\rm tNFW}(R) = \frac{m_{\textsc{\tiny NFW}}}{4\pi R (R+r_{\rm s})^2}\left(\frac{r_{\rm t}^2}{R^2+r_{\rm t}^2}\right),
\ee
which is also shown in Fig.~\ref{fig:cdm_density_profile}. Here, $R$ is the three-dimensional distance from the center of the subhalo and $r_{\rm t}$ is the tidal radius. Observe that for $R\gg r_{\rm t}$, the density profile decays quickly as $R^{-5}$. Basically, our truncation scheme is meant to reflect that any dark matter particles outside $r_{\rm t}$ are tidally stripped as the subhalo undergoes a full orbit within its host. The tidal radius thus evolves in time, generally getting smaller as the subhalo orbits within the tidal field of the host. 

Projecting Eq.~\eqref{eq:3DNFW} along the line of sight leads to the following convergence profile for a tNFW subhalo \cite{Baltz_2007} 
\begin{align}\label{eq:smd_tnfw}
\kappa_{\rm tNFW}(x) &= \frac{m_{\textsc{\tiny NFW}}}{ \Sigma_{\rm crit}r_{\rm s}^2} \frac{\tau^2}{2 \pi  (\tau^2 + 1)^2} \Bigg[\frac{\tau^2+1}{x^2-1}(1-F(x))   \en
& + 2F(x) - \frac{\pi}{\sqrt{\tau^2 + x^2}} + \frac{\tau^2-1}{\tau \sqrt{\tau^2 + x^2}} L(x) \Bigg],
\end{align}
where
\begin{align}
x=\frac{r}{r_{\rm s}},\quad \tau = \frac{r_{\rm t}}{r_{\rm s}},
\end{align}
\begin{align}
F(x) = \frac{\cos^{-1}(1/x)}{\sqrt{x^2-1}},
\end{align}
\begin{align}
L(x) = \ln\left(\frac{x}{\sqrt{\tau^2 + x^2} + \tau} \right).
\end{align}

The scale mass $m_{\textsc{\tiny NFW}}$ is related to the total  subhalo mass $m$ via the relation \cite{Baltz_2007}
\begin{align}
m = \frac{m_{\textsc{\tiny NFW}} \tau^2}{(\tau^2+1)^2}\left[(\tau^2-1)\ln(\tau) + \tau \pi - (\tau^2 + 1)\right].
\end{align}
The parameter $\tau$ is similar to the concentration parameter, $c_{\rm vir} = R_{\rm vir}/r_{\rm s}$, which measures how concentrated the mass of a halo is since most of the mass is contained within $r_{\rm s}$. The tidal radius and virial radius are not necessarily the same however, so $c_{\rm vir} \neq \tau$. 

In the notation of Sec. \ref{sec:subs_stats}, the internal structure parameters for a truncated NFW subhalo are simply $\qq = \{r_{\rm s}, r_{\rm t}\}$. Here, we adopt the following phenomenological relations between the internal structure parameters and the subhalo mass and position \cite{Cyr-Racine_2015}:
\begin{align}
r_{\rm s}  &= r_{{\rm s},0} \left(\frac{m}{m_0}\right)^{\gamma} \label{eq:rs_mass},\\
r_{\rm t} & =  r_{\rm t,0} \left(\frac{m}{m_0}\right)^{1/3}\left(\frac{r_{\rm 3D}}{r_{\rm 3D,0}}\right)^{\nu}\label{eq:rt_mass_pos},
\end{align}
where we adopt below a fiducial value of $\gamma = 1/3$ \citep{2014MNRAS.442.3598V,2008MNRAS.390L..64D}, and $\nu$ is a parameter that depends on the density profile of the host; for an isothermal profile $\nu = 2/3$, while $\nu = 1$ for a subhalo outside the scale radius of an NFW host \cite{Cyr-Racine_2015}. The quantity $r_{\rm 3D}$ is the three-dimensional distance between the subhalo and the center of the host galaxy, and $r_{\rm s, 0}$ and $r_{\rm t,0}$ are, respectively, the scale and truncation radii for a subhalo of mass $m_0$ at position $r_{\rm 3D,0}$. For a pivot mass $m_0 = 10^6$ $M_{\odot}$, we adopt $r_{\rm s,0} = 0.1$ kpc \citep{2014MNRAS.442.3598V}, $r_{\rm t,0} = 1$ kpc, and $r_{\rm 3D,0} = 100$ kpc \citep{Diemand:2008in,Springel_2008}.

In order to apply the result from the previous section, we need to know the distribution $\mathcal{P}_{\rm q}(r_{\rm s},r_{\rm t}| m)$, which we assume can be written as 
\be
\mathcal{P}_{\rm q}(r_{\rm s},r_{\rm t}| m) = \mathcal{P}_{\rm s}(r_{\rm s}| m) \, \mathcal{P}_{\rm t}(r_{\rm t}| m).
\ee
We model the distribution for scale radii assuming that the scatter in the scale radius-mass relation Eq.~\eqref{eq:rs_mass} is normally distributed such that
\be\label{eq:ps}
\mathcal{P}_{\rm s}(r_{\rm s}|m) = \mathcal{N}\left(r_{\rm s,0}\left(\frac{m}{m_0}\right)^{\gamma},\sigma_{r_{\rm s}} r_{\rm s,0} \left(\frac{m}{m_0}\right)^{\gamma}\right),
\ee
where $\mathcal{N}(\mu,\sigma)$ is a Gaussian probability distribution with mean $\mu$ and standard deviation $\sigma$, and $\sigma_{r_{\rm s}}$ is the fractional scatter about the scale radius-mass relation given in Eq.~\eqref{eq:rs_mass}. We take $\sigma_{r_{\rm s}} = 0.2$ throughout the rest of this paper, but we note that this specific choice has very little impact on our results. 

Noting that $r_{\rm 3D}^2 = r^2 + h^2$, where $h$ is the projection of $r_{\rm 3D}$ along the line of sight and $r$ is the projection onto the lens plane, the distribution of tidal radii marginalized over $h$ can be written as
\begin{align}\label{eq:integral_pt}
\mathcal{P}_{\rm t}(r_{\rm t}|m,r) &= \frac{1}{Z}  \int dh \; \mathcal{P}_{\rm 3D}\left(\sqrt{r^2+h^2}\right) \\
& \hspace*{1cm}\de\left( r_{\rm t} - r_{\rm t,0} \left(\frac{m}{m_0}\right)^{1/3}\left(\frac{\sqrt{r^2 + h^2}}{r_{\rm 3D,0}}\right)^{\nu}\right),\nonumber
\end{align}
where $\mathcal{P}_{\rm 3D}$ is the three-dimensional distribution of subhalos within the lens galaxy and $Z$ is a normalization factor equal to the projection integral,
\be
Z \equiv  \int dh \; \mathcal{P}_{\rm 3D}(\sqrt{r^2+h^2}) = \mathcal{P}_{\rm r}(r).
\ee

Under the assumption that the projected distribution of subhalos is uniform, the radial distribution of subhalos is simply equal to the inverse area of the strong lensing region, $\mathcal{P}_{\rm r} = 1/A$. The choice of $\mathcal{P}_{\rm 3D}$ to obtain this is not unique. However, in the limit that the strong lensing region is probing only a small projected area of the host lens galaxy, we can obtain a unique expression for $\mathcal{P}_{\rm t}$ even when the distribution of subhalos is nonuniform. As shown in Appendix \ref{sect:deriving_pt}, the integral in Eq.~\eqref{eq:integral_pt} can be performed in this limit to yield
\begin{align}\label{eq:general_pt}
\mathcal{P}_{\rm t}(r_{\rm t}|m) = \frac{1}{\nu R_{\rm max}} \frac{r_{\rm 3D,0}}{r_{\rm t}} \left[\left(\frac{m_0}{m}\right)^{1/3} \frac{r_{\rm t}}{r_{\rm t,0}} \right]^{1/\nu}.
\end{align}
In the following we model the host as being isothermal, for which $\nu = 2/3$ as stated above.
Note that $\mathcal{P}_{\rm t}$ has no dependence on the subhalo position within the host, consistent with the assumptions used in Sec. \ref{sec:subs_stats}.  

\begin{figure}[t!]
\includegraphics[width=0.49\textwidth]{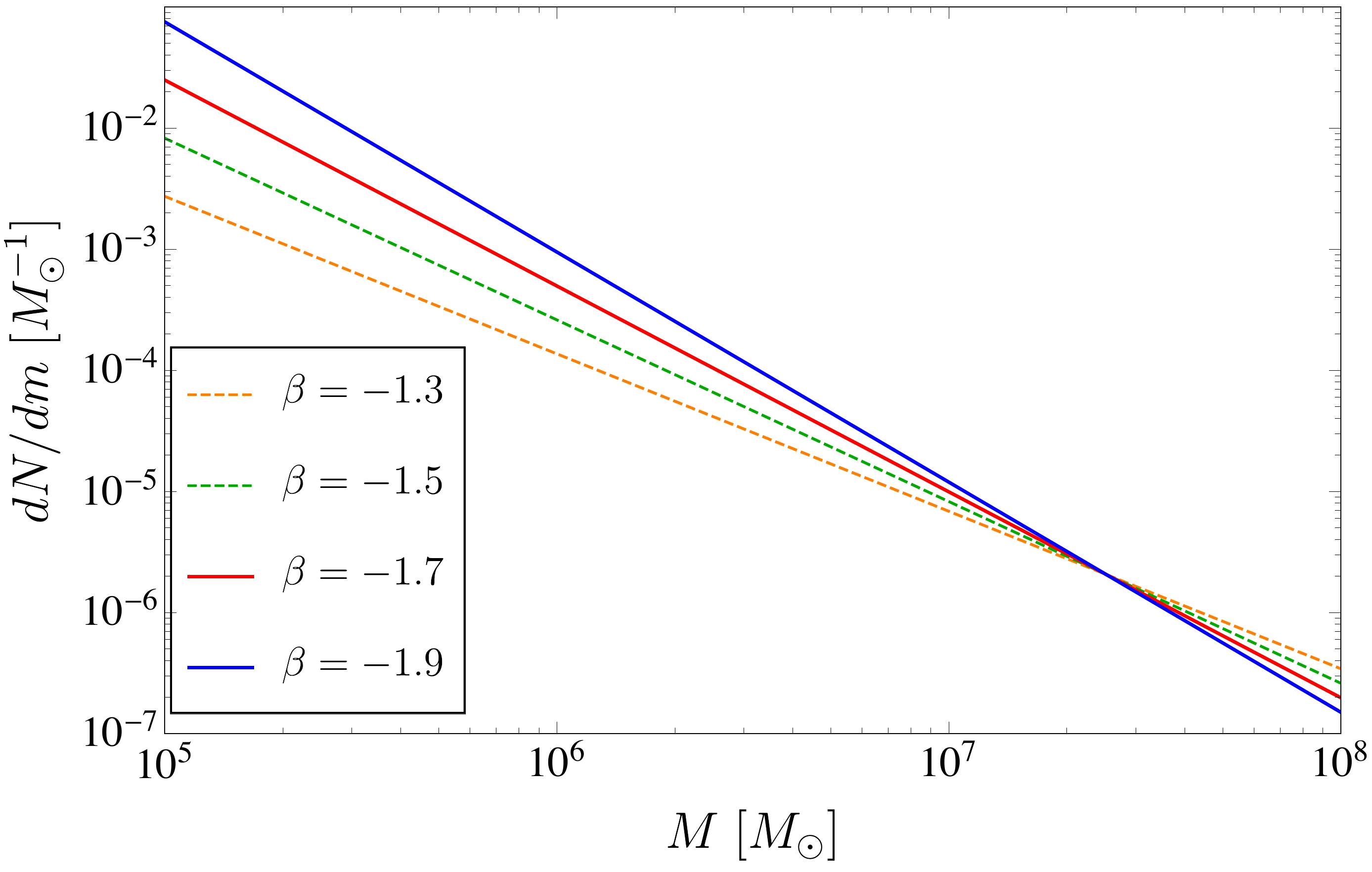}
\caption{Subhalo mass function (Eq. \ref{eq:dNdm}) for different values of the power-law index $\beta$.}\label{fig:dndm}
\end{figure}

Lastly, we express the mass probability distribution as a power-law function \cite{Springel_2008}
\begin{align}\label{eq:dNdm}
\frac{dN_{\rm sub}}{dm} = a_0 \left(\frac{m}{m_*}\right)^{\beta},
\end{align}
where $\beta = -1.9$ and $m_* = 2.52 \times 10^7$ $M_\odot$. This mass function is illustrated in Fig.~\ref{fig:dndm} for different choices of $\beta$. We note that the constant $a_0$, which normalizes the subhalo mass function, and the average convergence $\bar{\kappa}_{\rm sub}$ are proportional to one another as per Eq.~\eqref{eq:pract_to_phys}. Typical gravitational lenses have an average convergence in the range $0.003<\bksub<0.03$  \cite{Dalal:2002aa}, so we normalize the subhalo mass function such that $\bksub = 0.02$. 

Although we do not require the convergence field to be Gaussian, nor do we assume it, we do limit the large non-Gaussian contributions from the few most massive subhalos by setting an appropriate upper bound on the subhalo mass range included in our analysis. In practice, the maximum subhalo mass to include in the substructure convergence power spectrum calculation should be dictated by the data set used to measure it. Indeed, the spatial resolution, pixel size, and the signal-to-noise ratio of the data specifies a subhalo mass sensitivity threshold below which a statistically significant direct detection of a subhalo is unlikely. For high-quality space-based optical data, this threshold could be as low as $\sim 10^8$ $M_\odot$ \citep{Vegetti2014}, while for interferometric data it could reach $\sim10^7$ $M_\odot$ \citep{Hezaveh_2013}. Here, we adopt a fiducial value of $m_{\rm high} = 10^8$ $M_{\odot}$. The minimum subhalo mass we consider is $m_{\rm low} = 10^5$ $M_{\odot}$. As we will show below, the specific choice of $m_{\rm low}$ is largely inconsequential as long as $m_{\rm low} \ll m_{\rm high}$. 

%%%%%%%%%%%%%%%%%%%%%%%%%%%%%%%%%%%%%%%%%%%%%%%%%%%%
\subsection{Power spectrum: 1-subhalo term}

We can now apply the formalism developed in Sec. II to a population of tNFW subhalos to study how the abundance, density profile, radial distribution, and subhalo sizes affect the the convergence power spectrum. In this case, Eq.~\eqref{eq:1sh_constdist} for the 1-subhalo term becomes
\begin{align}
P_{\rm 1sh}(k) &=\frac{ \bksub}{\langle m\rangle \Sigma_{\rm crit}} \int dm \; m^2 \; \mathcal{P}_{\rm m}(m) \int dr_{\rm s} \; dr_{\rm t} \; \mathcal{P}_{\rm s}(r_{\rm s}|m) \en
&\hspace*{1.5cm}\times \mathcal{P}_{\rm t}(r_{\rm t}|m) \, |\tilde{\kappa}(k,r_{\rm s},r_{\rm t})|^2.
\end{align}

%%%%%%%%%%%%%%%%%%%%%%%%%%%%%%%%%%%%%%%%%%%%%%%%%%%%
\subsubsection{Analytical discussion}\label{sec:analytical_disc}

For typical Poisson realizations of a population of spherically symmetric tNFW subhalos, we expect the behavior of the 1-subhalo term to depend mostly on three quantities: a low-$k$ power spectrum amplitude, a turnover scale $k_{\rm trunc}$ corresponding approximately to the size of the largest subhalos, and an asymptotic high-$k$ slope dictated by the small-$r$ behavior of the subhalo density profile, which takes over for $k\gg k_{\rm scale}$ (defined below). 

For small $k$ values, we expect the 1-subhalo contribution to the power spectrum to plateau to a constant value since taking $k \rightarrow 0$ makes $J_0(kr) \rightarrow 1$ in Eq.~\eqref{eq:1sh_constdist}, in which case $\kappa$ and $P_{\rm 1sh}$ are $k$-independent. Another way to understand this low-$k$ plateau is to realize that subhalos can be modeled as point masses, i.e.~$\hat{\kappa}_i =\delta^{(2)}(\rr - \rr_i)$, on scales larger than the biggest subhalo's truncation radius, hence leading to $P_{\rm1sh}(k)= \bksub \langle m^2\rangle/(\langle m \rangle\Sigma_{\rm crit})$. With $\bksub = 0.02$ and our choice for the mass function parameters described above, we expect a low-$k$ amplitude of $\sim 10^{-4}$ kpc$^2$. 

As $k$ is increased, the power spectrum begins probing the actual density profile of the subhalos, leading to a suppression of the power compared to the pure point-mass case. This turnover scale is determined by the truncated size of the largest subhalos, since this is the largest scale in the problem relevant to the 1-subhalo term. We therefore expect that this turnover is going to occur near a scale that corresponds to the inverse of the tidal radius of the largest subhalo: $k_{\rm trunc} \equiv 1/r_{\rm t,max}$.  

As $k$ is further increased, the 1-subhalo term probes the intermediate scales between the typical truncation and scale radii of the tNFW subhalo population. Finally, we expect the convergence power spectrum to asymptote to a power-law behavior at large $k$ where it is probing scales deep within the NFW scale radius. This power law can be determined by finding the small-$x$ limit of the convergence profile given in Eq.~\eqref{eq:smd_tnfw},
\begin{align} \label{eq:lim}
\kappa_{\rm tNFW}(x) \xrightarrow{} \frac{m_{\tiny\rm NFW}}{2 \pi r_{\rm s}^2\Sigma_{\rm crit}} \left( \ln\left(\frac{2}{x}\right) -1\right),\quad x\ll1,
\end{align}
and taking the (2D) Fourier transform, which leads to 
\be
\tilde{\kappa}_{\rm tNFW}(k) \xrightarrow{}  \frac{1}{(k \,r_{\rm s})^2}, \quad k r_{\rm s} \gg 1. 
\ee
This implies that $P_{\rm1sh}(k)\propto 1/k^4$ for $k r_{\rm s}\gg 1$. We expect the power spectrum to reach this slope at a scale below that of the smallest scale radii in the population. It is therefore useful to define the wave number $k_{\rm scale} \equiv 1/r_{\rm s,min}$ beyond which the convergence power spectrum is a simple power law determined by the inner density profile of the subhalos. 

%%%%%%%%%%%%%%%%%%%%%%%%%%%%%%%%%%%%%%%%%%%%%%%%%%%%
\subsubsection{Numerical results}

\begin{figure*}[t!]
	\centering
	\begin{subfigure}[t]{0.497\textwidth}
		\centering
		\includegraphics[width = \textwidth]{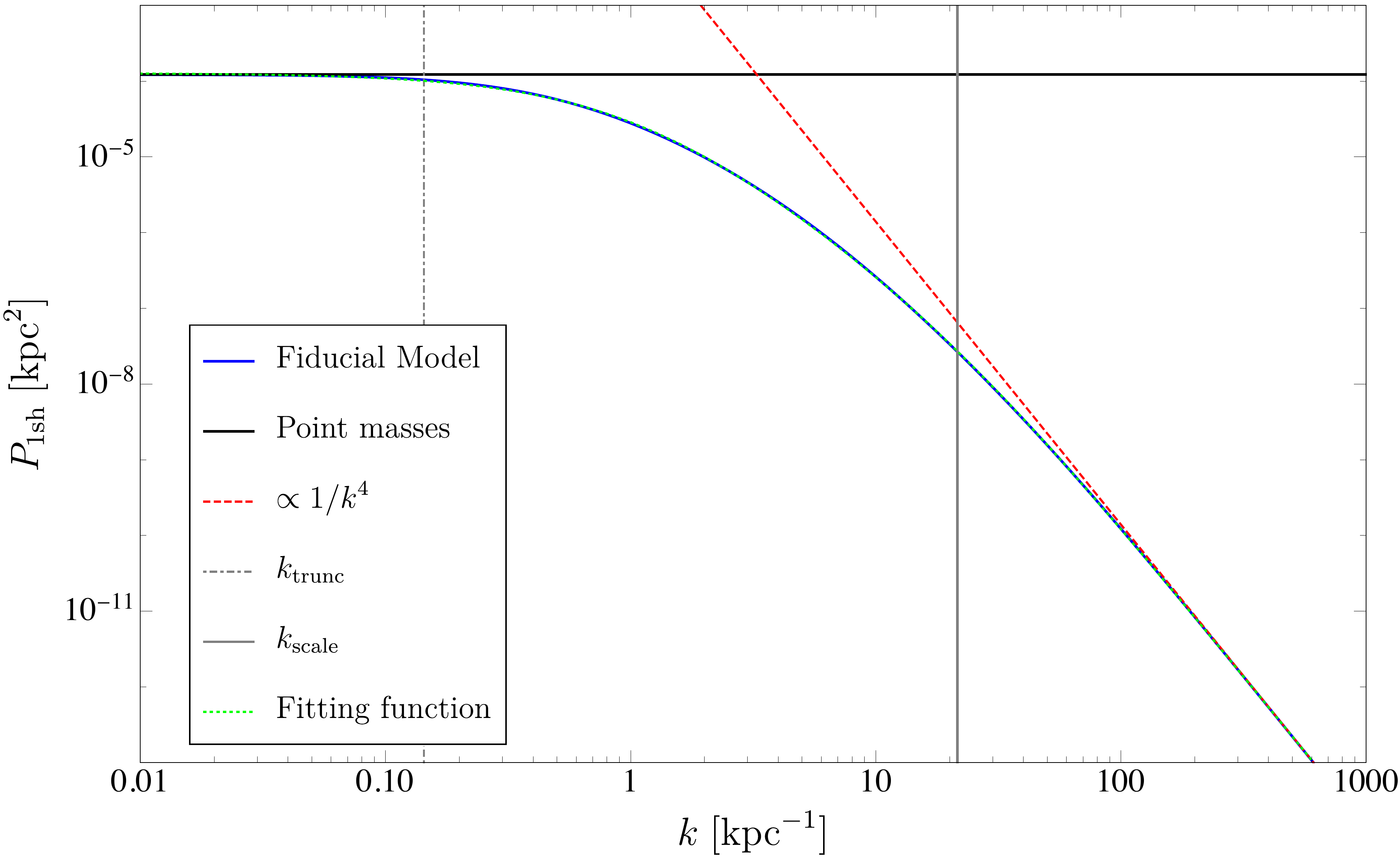}	
		\caption{}
		\label{fig3:Panel_a}
	\end{subfigure}
	\begin{subfigure}[t]{0.497\textwidth}
		\centering
		\includegraphics[width = \textwidth]{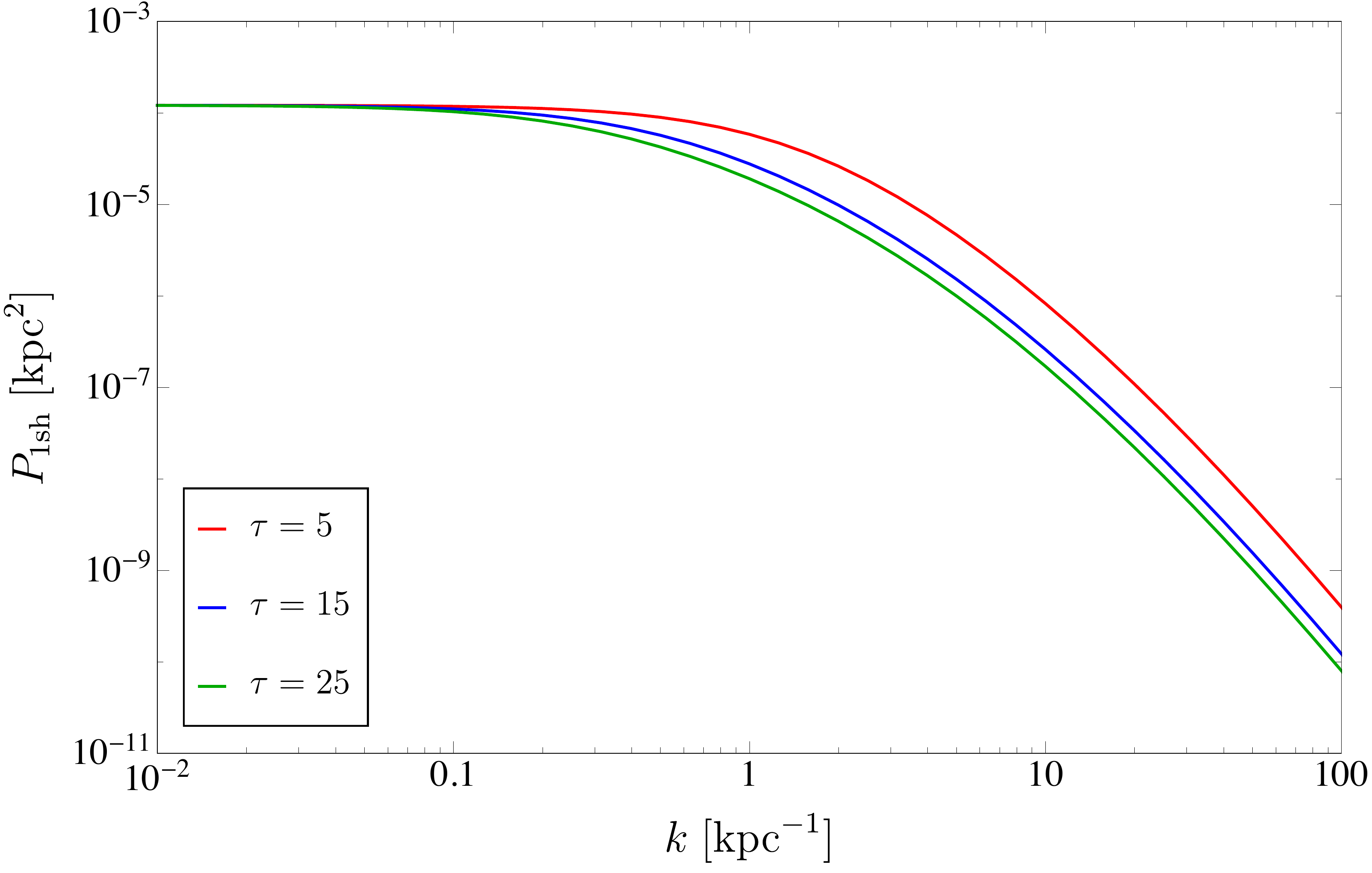}
		\caption{}
		\label{fig3:Panel_b}
	\end{subfigure}
	\\
	\begin{subfigure}[t]{0.497\textwidth}
		\centering
		\includegraphics[width = \textwidth]{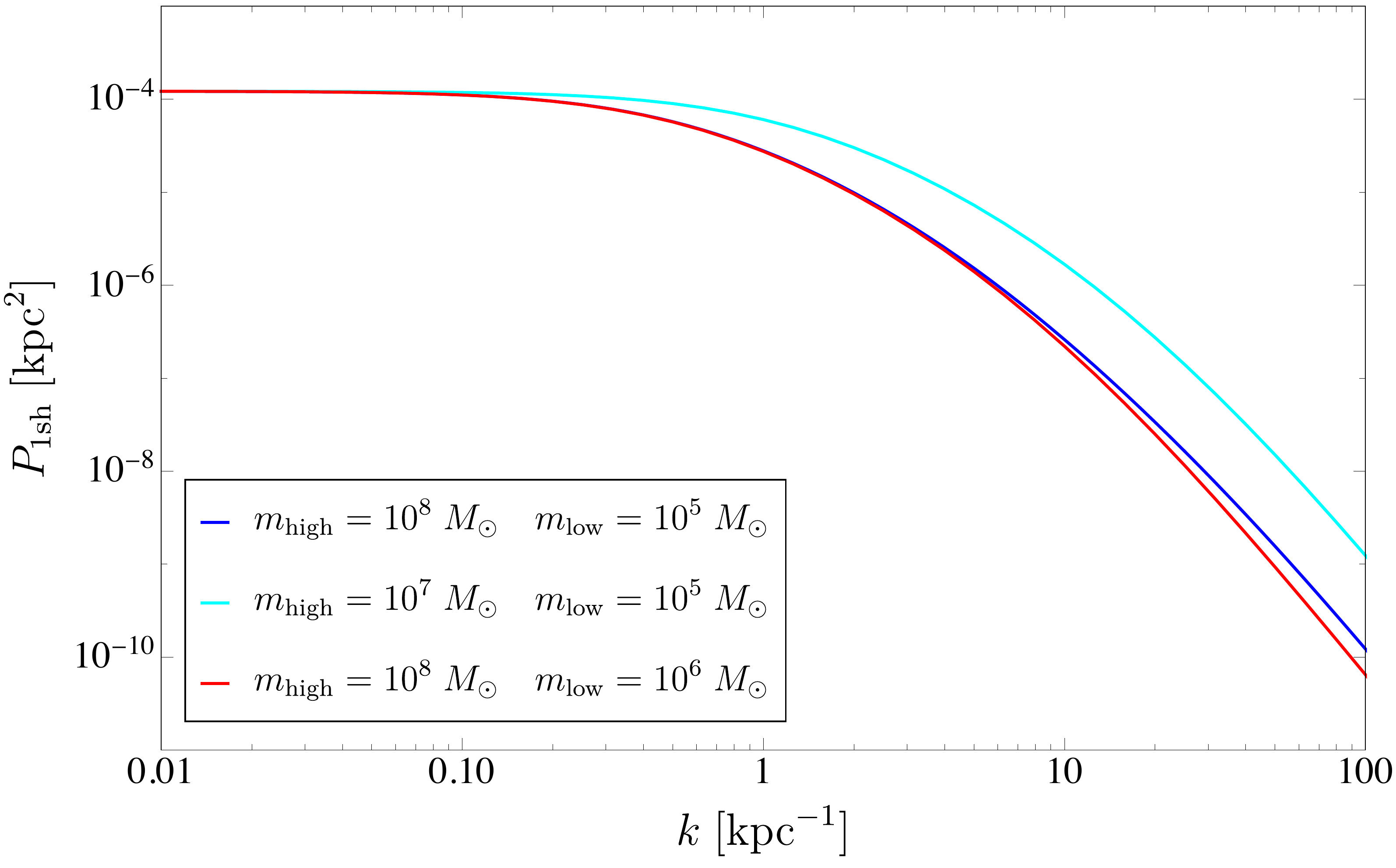}
		\caption{}
		\label{fig3:Panel_c}
	\end{subfigure}
	\begin{subfigure}[t]{0.497\textwidth}
		\centering
		\includegraphics[width = \textwidth]{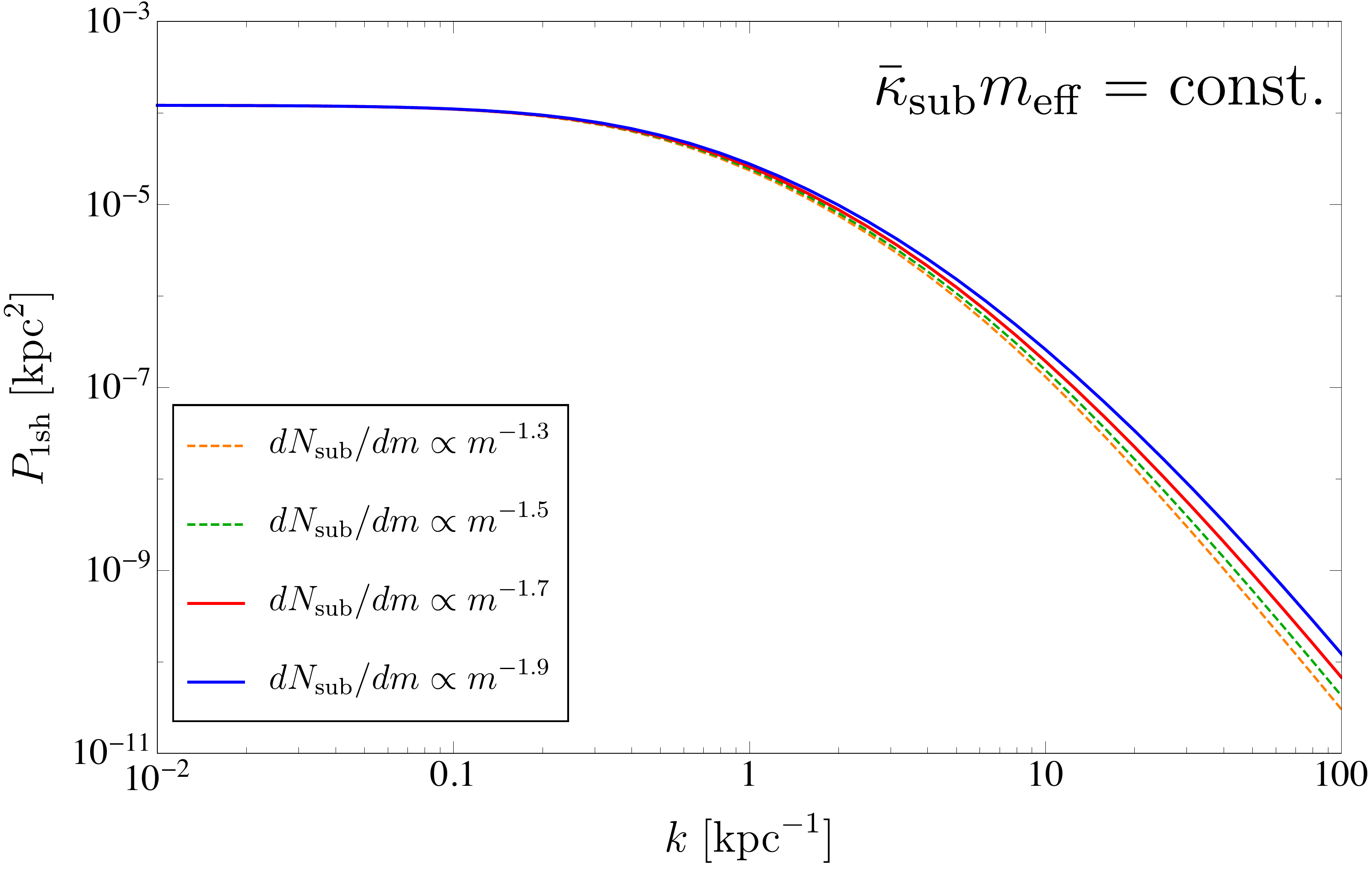}
		\caption{}
		\label{fig3:Panel_d}
	\end{subfigure}
\caption{The 1-subhalo term of the convergence power spectrum of a population of truncated NFW subhalos. The solid blue line that appears in every subfigure represents the fiducial model with $\tau = 15$, $10^5$ M$_{\odot} \leq m \leq  10^8$ $M_{\odot}$, $r_{\rm s}$ given by Eq.~\eqref{eq:rs_mass} with $\gamma = 1/3$, and $dN_{\rm sub}/dm$ given by Eq.~\eqref{eq:dNdm} with $\beta = -1.9$. Panel (a) shows the features outlined in Sec. \ref{sec:analytical_disc}: the low-$k$ amplitude of the power spectrum matches that of a population of point masses (solid black); the high-$k$ slope is proportional to $1/k^4$ (dashed red); $k_{\rm trunc} \equiv 1/r_{\rm t,max} = 0.14$ kpc$^{-1}$ (dotted-dashed gray); and $k_{\rm scale} \equiv 1/r_{\rm s,min} = 21.5$ kpc$^{-1}$ (solid gray). The dotted green line corresponds to the fitting function described by Eqs.~(51) - (56). In Panels (b) - (d) we change one parameter in the fiducial model while leaving the others unchanged. (b): changing $\tau$ by keeping $r_{\rm s}$ unchanged but increasing $r_{\rm t}$. (c): decreasing (increasing) $m_{\rm high}$ ($m_{\rm low}$) by an order of magnitude. (d): decreasing the slope of the mass function down to $\beta= -1.3$. In Panels (c) and (d) $\bksub m_{\rm eff}$ is held constant as the parameters are varied, where $m_{\rm eff} \equiv \langle m^2\rangle/\langle m\rangle$. Note the different horizontal axis in Panel (a) and Panels (b) - (d). }\label{fig:powerspec_hezaveh}
\end{figure*}
Before ensemble-averaging over $\mathcal{P}_{\rm s}$ and $\mathcal{P}_{\rm t}$, it is informative to consider the shape of the convergence power spectrum for specific values of $r_{\rm s}(m)$ and $r_{\rm t}(m)$. Making the following choices:
 \begin{align}
 \mathcal{P}_{\rm s}(r_{\rm s}|m) = \delta \left(r_{\rm s} - r_{\rm s,0}\left(\frac{m}{10^9 \; M_{\odot}}\right)^{\gamma}\right),
 \end{align}
 \begin{align}\label{eq:pt_deltafunc}
 \mathcal{P}_{\rm t}(r_{\rm t}|m) = \delta(r_{\rm t} - 15r_{\rm s}),
 \end{align}
the 1-subhalo term takes the simple form
\begin{align}\label{eq:ps_hezaveh}
P_{\rm 1sh}(k) &=\frac{ \bksub}{\langle m\rangle \Sigma_{\rm crit}} \int dm \; m^2 \;\mathcal{P}_{\rm m}(m) \;|\tilde{\kappa}(k,m)|^2. 
\end{align}
Note that Eq.~\eqref{eq:pt_deltafunc} is equivalent to having a constant ratio for $\tau = r_{\rm t}/r_{\rm s} = 15$, which is not generally the case. From our expressions for the scale and tidal radius we expect $\tau$ to lie in the range $ \approx 1 - 25$, depending on subhalo mass and position. 

Figure 3 ~\ref{fig:powerspec_hezaveh} shows the power spectrum defined in Eq.~\eqref{eq:ps_hezaveh}. Panel (a) displays the features discussed in the preceding section, which have the expected behavior. The asymptotic low-$k$ amplitude is $1.2 \times 10^{-4}$ kpc$^2$ and matches the amplitude of the power spectrum of a population of point masses (black) with the same mass function. The truncation scale, which for $r_{\rm t,max} \simeq 7$ kpc is $k_{\rm trunc} = 0.14$ kpc$^{-1}$ (dashed-dotted gray), very closely matches the scale at which the power spectrum turns over, consistent with the fact that this scale corresponds to the sizes of the largest subhalos. Furthermore, past $k_{\rm scale} = 21.5$ kpc$^{-1}$ (gray) the large-$k$ behavior matches a power law $1/k^4$ (dashed red), which again matches our expectation since in this regime we are within the scale radius of even the smallest subhalos i.e., where the tNFW convergence goes as Eq.~\eqref{eq:lim}.

In the remaining panels we vary several parameters of relevance to the power spectrum. Panel (b) shows the effect of changing the density profiles of subhalos by changing $\tau$. When we increase $\tau$, we are keeping $r_{\rm s}$ and $m$ fixed while increasing $r_{\rm t}$, which means that the subhalo size is increasing and subhalos are becoming less concentrated toward the center. This has the effect of decreasing power on small scales and decreasing $k_{\rm trunc}$. 

Panels (c) and (d) both reflect changes in the subhalo mass function: the former shows the result of varying $m_{\rm high}$ and $m_{\rm low}$, and the latter, the effect of making the power law shallower. Both changes affect the low-$k$ amplitude as well as the distribution of power and slope on scales larger than $k_{\rm trunc}$; to disentangle these two effects we keep the quantity $\bksub m_{\rm eff} = \bksub \langle m^2\rangle/\langle m\rangle$ fixed while changing the mass function, which makes the low-$k$ amplitude remain the same. In this manner, we can isolate the effects of the subhalo mass function on the shape of the convergence power spectrum at high $k$. In Panel (c) we see that decreasing $m_{\rm high}$ by an order of magnitude adds power on small scales. Indeed, removing the largest subhalos and redistributing their mass among smaller subhalos causes an increase in $k_{\rm trunc}$, which adds power on small scales. Panel (c) also illustrates the impact of increasing $m_{\rm low}$ from $10^5$$M_\odot$ to $10^6$$M_\odot$. The resulting change to the convergence power spectrum is rather small, reflecting the fact that the more massive subhalos tend to dominate the behavior of the power spectrum. This also implies that the convergence power spectrum shows little sensitivity to the low-mass cutoff of the mass function. Finally, Panel (d) shows that, by making the power law shallower, we are reducing power on small scales. To understand this effect, we refer the reader to Fig.~\ref{fig:dndm}, where one can see that by making the slope shallower, we are decreasing the number of low-mass subhalos and are in fact increasing the number of subhalos more massive than the pivot mass. Note that despite the change in the shape of the power spectrum on intermediate scales, the spectra still match the $1/k^4$ power law of the fiducial case at $k \gtrsim k_{\rm scale}$.

\begin{figure*}[t!]
\centering
	\begin{subfigure}[t]{0.497\textwidth}
		\centering
		\includegraphics[width=\textwidth]{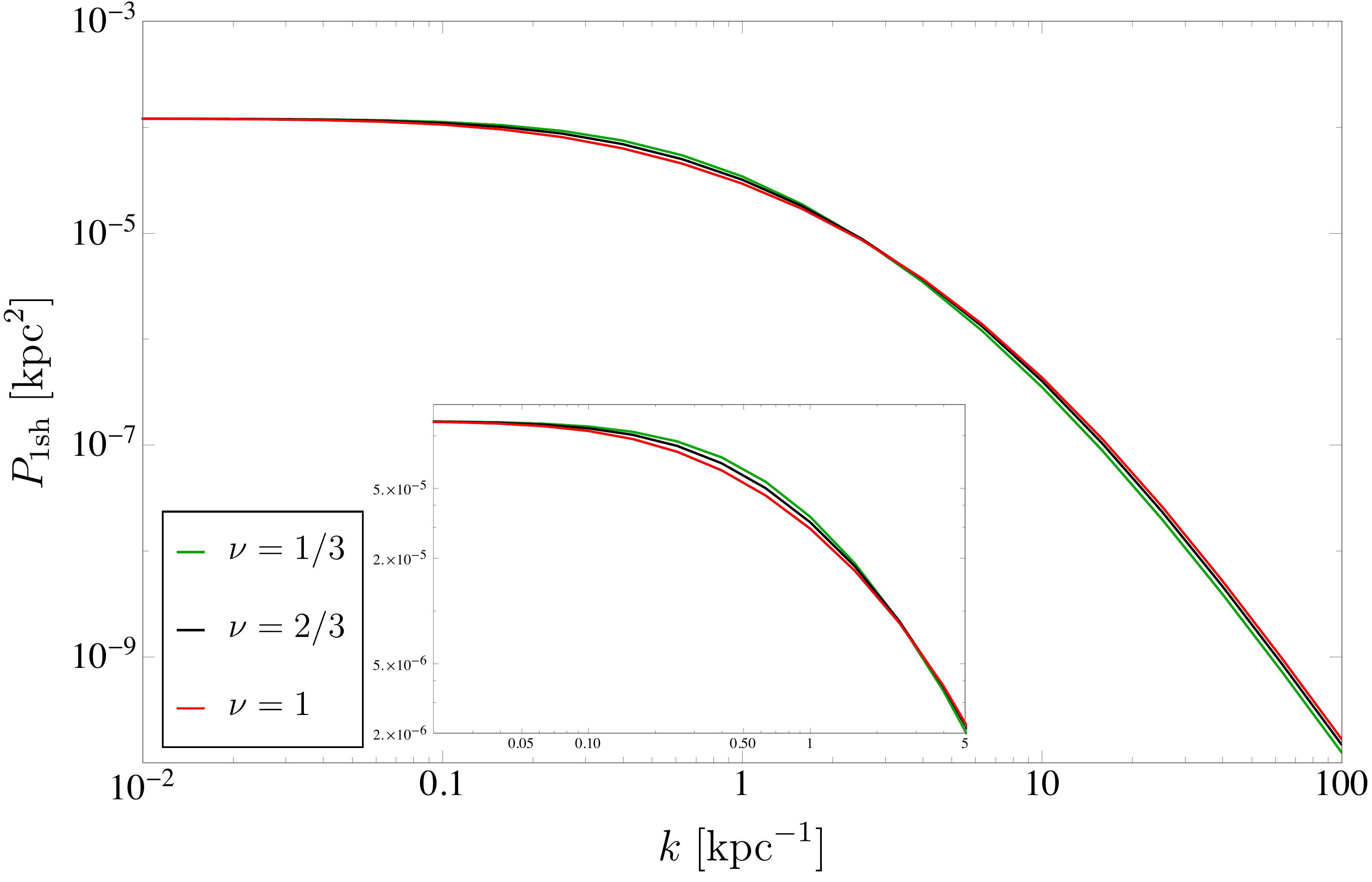} 
		\caption{}
		\label{}
	\end{subfigure}%
	\begin{subfigure}[t]{0.497\textwidth}
		\centering
		\includegraphics[width=\textwidth]{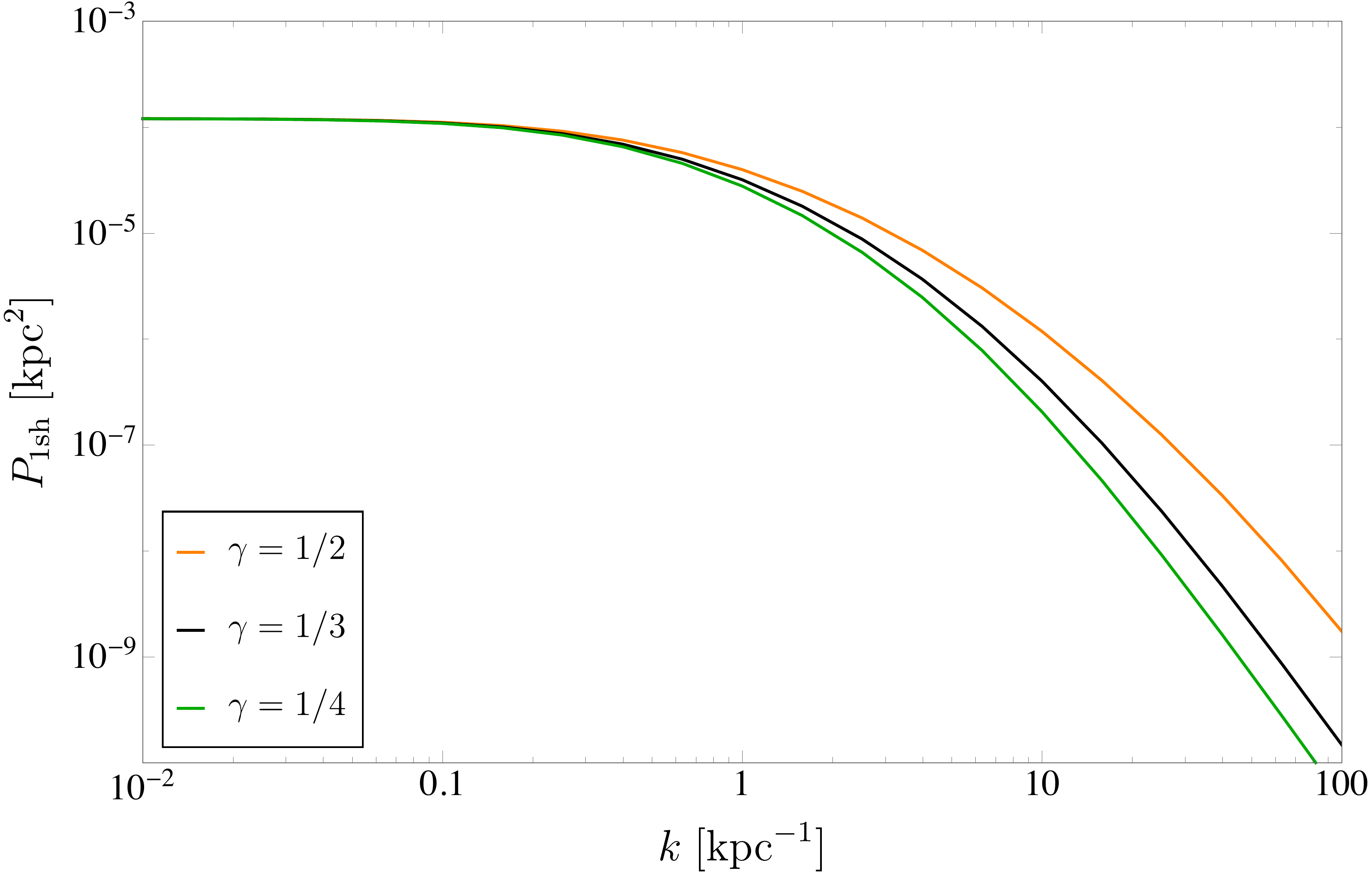}
		\caption{}
		\label{}
	\end{subfigure}%
\caption{ Ensemble-averaged 1-subhalo term for a population of truncated NFW halos. The black line that appears in both panels has parameter values equal to the fiducial model in Fig. \ref{fig:powerspec_hezaveh} (except for $\tau$, which we do not fix). There are two additional parameters: $\nu = 2/3$ and $\sigma_{r_{\rm s}}=0.2$. Panel (a) varies the power-law dependence of the tidal radius on $r_{\rm 3D}$, Eq.~\eqref{eq:rt_mass_pos}. Panel (b) varies the power law of the scale radius-mass relation, Eq.~\eqref{eq:rs_mass}. }\label{fig:p1sh_ensavg}
\end{figure*}

Having gained some intuition into how different parameters in our model affect the power spectrum, we can  move on to the more general case where we perform ensemble averages over the two intrinsic subhalo parameters: $r_{\rm s}$ and $r_{\rm t}$. The 1-subhalo power spectrum in this case is shown in Fig.~\ref{fig:p1sh_ensavg}. The fiducial model -- shown in black in both panels -- corresponds to the parameter values for $\mathcal{P}_{\rm t}$ and $\mathcal{P}_{\rm s}$, given in Eqs.~\eqref{eq:ps} and \eqref{eq:general_pt}), $\nu = 2/3$ (isothermal lens) and $\gamma =1/3$. 

In each panel we show the effect of changing one of these parameters. Panels (a) and (b) reflect changes in $\nu$ and $\gamma$, respectively. It is immediately obvious from Panel (a) that changing the index $\nu$ has little impact on the convergence power spectrum, beside from a slight redistribution of power at intermediate and small scales. This means that the power spectrum will have limited sensitivity to the host galaxy's density profile; on the other hand, it also means that uncertainties on the density profile of the host will not prevent the power spectrum from being an effective tool to study subhalo populations. 

Panel (b) of Fig.~\ref{fig:p1sh_ensavg} demonstrates that the power law in the scale radius-mass relation can have a significant impact on the small-scale substructure convergence power spectrum. As we increase $\gamma$, the minimum scale radius decreases quickly, and so $k_{\rm scale}$ increases. In fact $r_{\rm s,min}$ decreases by an order of magnitude as we change $\gamma$ from 1/4 to 1/2. This has the effect of adding  power on small scales, as discussed in Sec. \ref{sec:analytical_disc}. 

Another natural parameter to vary would be the scatter in the scale radius-mass relationship, $\sigma_{r_{\rm s}}$. However, for a scatter of $20\%$ or less, the impact on the convergence power spectrum is much smaller than the change associated with varying the index $\gamma$, and we therefore do not show it here. We also note that for a scatter larger than $\sim 20\%$, the approximate model presented in Eq.~\eqref{eq:ps} likely breaks down at small subhalo masses, and should be replaced by a more realistic distribution of $\mathcal{P}_{\rm s}(r_{\rm s}|m)$.

We find that the 1-subhalo term for a population of tNFW halos is well fit by a function of the form
\begin{align}
P_{\rm 1sh}(k) &= \frac{g_0}{1+g_1 k+\left( g_2 k\right)^2+\left(g_3 k\right)^3+\left(g_4 k\right)^4},
\end{align}
where 
\begin{align}
g_0 = \frac{\bksub \langle m^2 \rangle}{\Sigma_{\rm crit}\langle m \rangle},
\end{align}
\begin{align}
g_1 = \frac{(1/3)}{\gamma}\frac{\langle \tau \rangle r_{\rm s,max}}{2 \pi},
\end{align}
\begin{align}
g_2 = \left(\frac{(1/3)}{\gamma}\right)^2 \frac{\langle \tau \rangle r_{\rm s,max}}{2 \pi},
\end{align}
\begin{align}
g_3 = r_{\rm s,max},
\end{align}
\begin{align}
g_4 &= \frac{\langle m^2 \rangle}{\int  \frac{dm \, d r_{\rm t} \,dr_{\rm s}\, m^2 \mathcal{P}_{\rm m}(m)\,\mathcal{P}_{\rm s}(r_{\rm s}|m) \,\mathcal{P}_{\rm t}(r_{\rm t}|m)}{r_{\rm s}^4 \left( \frac{\tau^2}{(\tau^2+1)^2}\left[(\tau^2-1)\ln(\tau) + \tau \pi - (\tau^2 + 1)\right] \right)^{2}}}. 
\end{align}
As shown, the parameters $g_i$ are determined by the truncation, the scale radius, the mass function, and the mass-concentration relation. We note that this fit works best for values of $\gamma \leq 1/3$, and starts deviating from the ``true" curve for higher values of $\gamma$. In the above, we have defined
\be
\langle \tau \rangle \equiv \int dm\, d r_{\rm t}\, dr_{\rm s} \mathcal{P}_{\rm m}(m)\,\mathcal{P}_{\rm s}(r_{\rm s}|m) \,\mathcal{P}_{\rm t}(r_{\rm t}|m)\frac{r_{\rm t}}{r_{\rm s}}.
\ee
The fitting function is shown as a dotted green line in Panel (a) of Fig.~\ref{fig:powerspec_hezaveh}.

%%%%%%%%%%%%%%%%%%%%%%%%%%%%%%%%%%%%%%%%%%%%%%%%%%%%
\subsection{Power spectrum: 2-subhalo term}

To find the total power spectrum we have to include the contribution of the 2-subhalo term, given by Eq.~\eqref{eq:p2sh}. As explained in Ref.~\cite{Chamberlain:2014}, the 2-subhalo term receives contributions from two distinct effects. First, subhalos have, in general, a nonuniform spatial distribution ($\mathcal{P}_{\rm r}(\rr)$ from Eq.~\eqref{eq:p_of_r}) due to their interaction with the potential well of their host halo. This so-called ``host'' contribution simply reflects the fact that subhalos can be gravitationally bound to their host lens galaxy, hence leading to a local enhancement of the convergence's two-point function. Second, subhalos can form self-bound groups orbiting their host galaxy. Due to tidal interactions with the latter, however, these subhalo groups are not expected to survive for more than a few dynamical times, \cite{Chamberlain:2014} and we thus foresee their contribution to be subdominant. So far, this contribution to $\xi_{\rm ss}(\rr)$ has not been measured nor extracted from simulations, at least at the mass scale of interest (see Ref.~\cite{Fang:2016usm} for a measurement on cluster scales.). Due to this, we focus below on the host contribution, but the reader should keep in mind that the subhalo group contribution should be added in order to get a fully accurate estimate of the 2-subhalo term.

\begin{figure}[t!]
\includegraphics[width=0.49\textwidth]{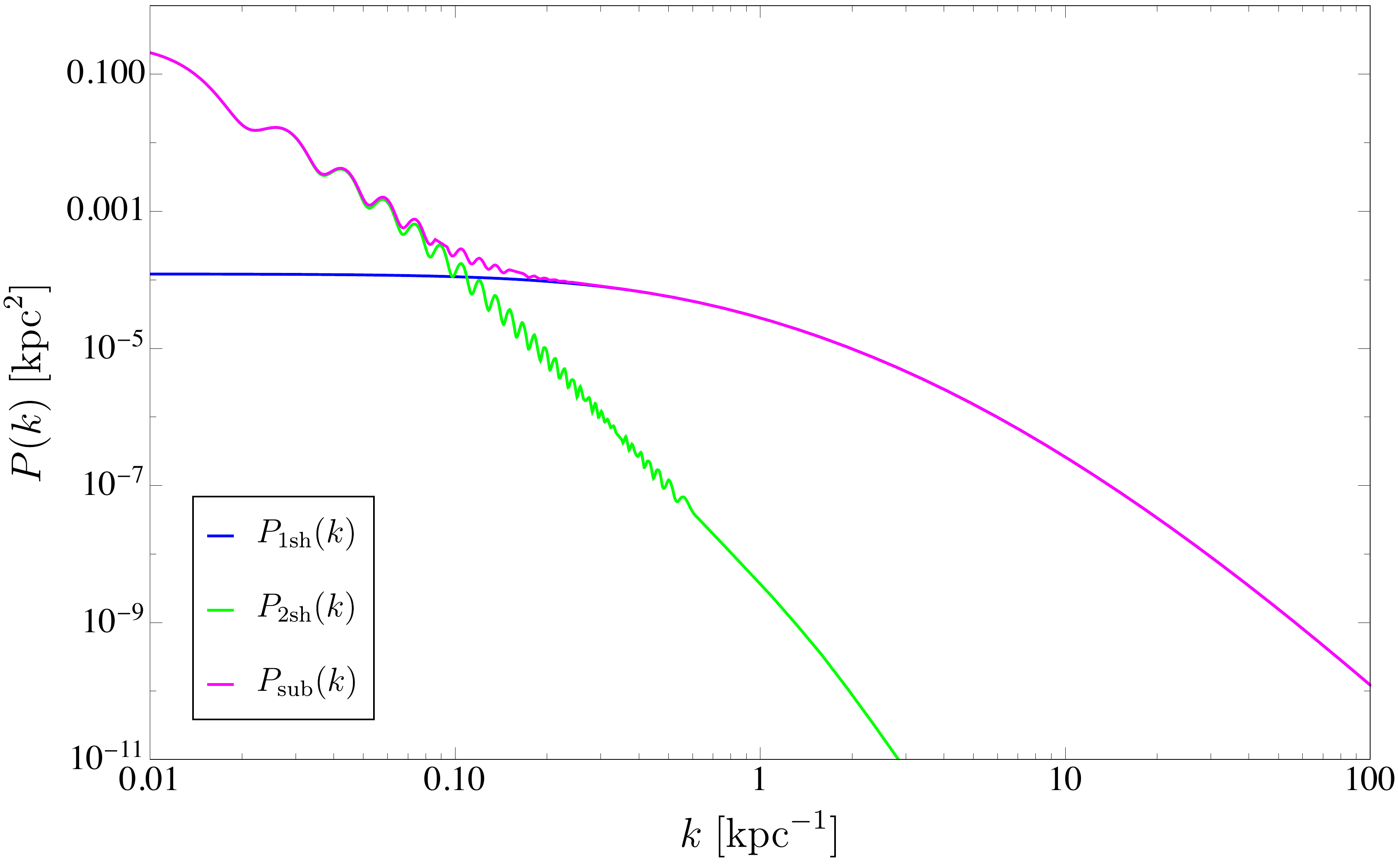}
\caption{Full convergence power spectrum (magenta) and individual contributions from the 1-subhalo (blue) and 2-subhalo (green) terms, where the radial subhalo distribution used to calculate the 2-subhalo term is given by Eq.~\eqref{eq:cored_pr}. }\label{fig:powerspec_total}
\end{figure}

As an illustrative example, we choose a radial distribution of subhalos that is cored and decays as $1/r$ for large $r$,
\begin{equation}\label{eq:cored_pr}
\mathcal{P}_{\rm r}(r) = \frac{1}{2 \pi (a + r) \left(R_{\rm max} + a \log\left(\frac{a}{a+R_{\rm max}}\right)\right)},
\end{equation}
where $a = 10$ kpc correponds to the core size. The total power spectrum $P_{\rm sub}(k)$ is shown in Fig.~\ref{fig:powerspec_total}, together with the individual contribution of the 1- and 2-subhalo terms. On large scales, for $k \lesssim 0.1$ kpc$^{-1} = 1/a$, the 2-subhalo term dominates, adding power and changing the low-$k$ slope from a constant to a power law. On small scales, however, the 1-subhalo term dominates (as expected), and the addition of the 2-subhalo term leaves the power spectrum unchanged. Note that the oscillations at small $k$ come from having $\mathcal{P}_{\rm r}(r)$ nonzero over a finite region in the lens plane.

\bigskip

%%%%%%%%%%%%%%%%%%%%%%%%%%%%%%%%%%%%%%%
\section{Truncated Cored subhalo population}
\label{sec:sidm}

In Sec. \ref{sec:subs_stats} we applied the convergence power spectrum formalism to a population of truncated NFW subhalos, since CDM halos in simulations seem to universally have NFW density profiles. We now apply the same methodology to a population of subhalos whose density profiles approximate what we expect SIDM subhalos to look like: cored at the center and with a large-$r$ behavior similar to NFW. The idea is to gauge the extent to which the power spectrum differs for NFW and cored profiles, which could be indicative of the utility of this observable in discerning between CDM and a different dark matter scenario in which halos are predicted to have cores instead, like SIDM. As we have emphasized in preceding sections, there are essentially two types of ingredients that go into the convergence power spectrum: the statistical properties of the subhalo population and the internal subhalo parameters, which determine the surface mass density profile. 

With respect to the first point, SIDM N-body simulations have shown that, at least in the case of elastic scattering with cross section $\sigma/m\lesssim1$ cm$^2/$g, the spatial distribution and number density of subhalos are largely unchanged \cite{Vogels_2012,Rocha_2012,Peter:2012jh,Zavala_2013}. Indeed, we expect that the subhalo distribution on the lens plane will be largely intact with respect to the CDM case since the volume occupied from the outskirts of the lens galaxy to the edge of its central region, where dark matter self-interactions can play a role, is many orders of magnitude larger than the volume occupied by the host's core itself; in fact the latter makes up about $\sim 2\%$ of the total line-of-sight volume. Furthermore, simulations find that there is essentially no change to the subhalo mass function for moderate dark matter self-interaction cross sections (at least down to $10^6$ $M_{\odot}$; refer to Fig.~6 of Ref.~\cite{Vogels_2012} to see both of these points). 

\begin{figure}[t]
\includegraphics[width=0.495\textwidth]{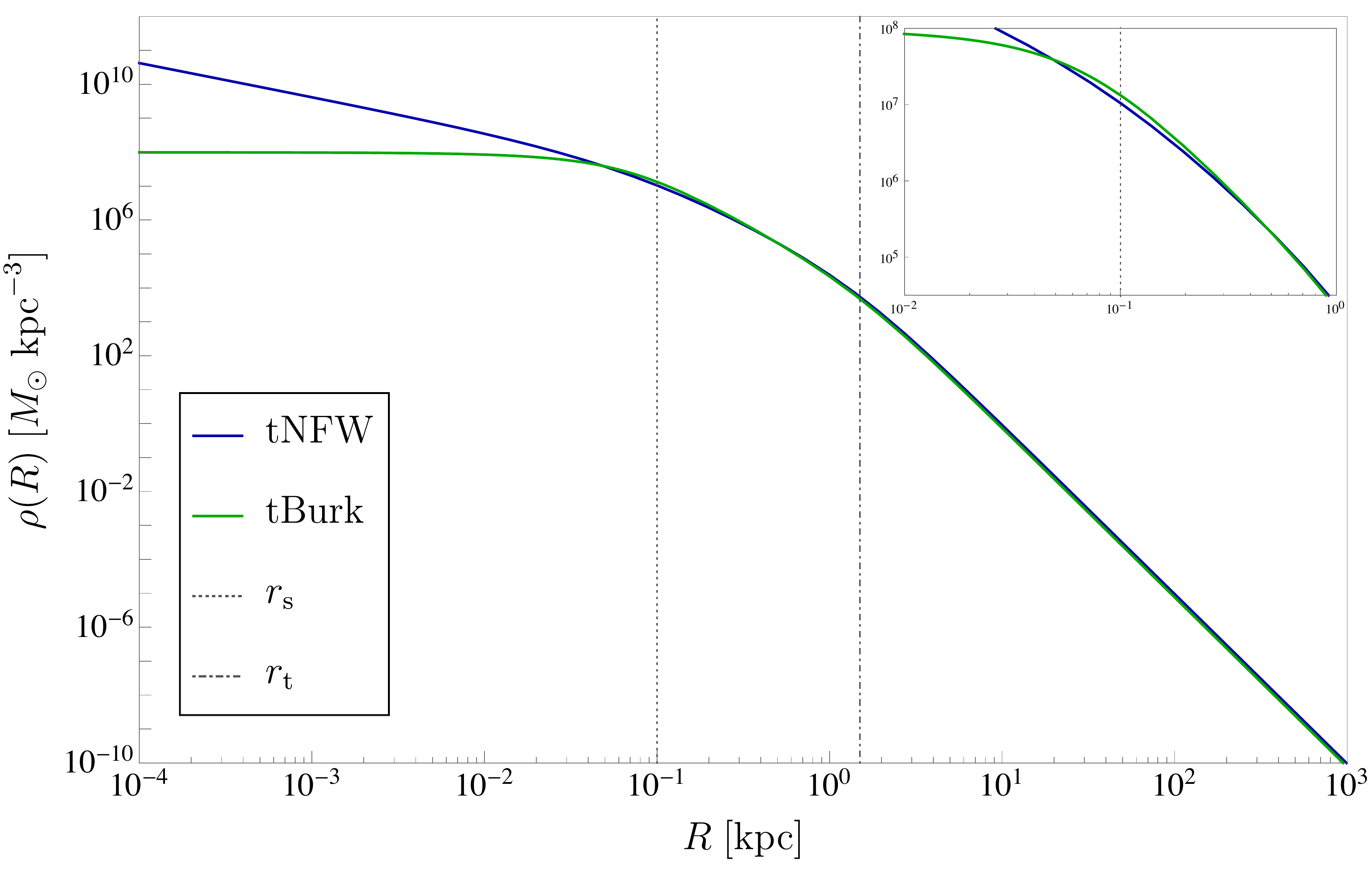}
\caption{Density profile for a truncated NFW profile (solid blue) and a truncated Burkert profile (solid green) for $\tau = 15$, $p=0.7$, and $m = 10^6$ $M_{\odot}$. The gray dotted and dashed-dotted lines represent the scale and tidal radius, respectively.}\label{fig:sidm_density_profile}
\end{figure}

With respect to the second point, there is a stark contrast between CDM and SIDM dark matter halos due to the appearance of a central core in the latter. A common cored density profile is the Burkert profile \cite{Burkert_1995},
\begin{equation}\label{eq:burkert}
\rho_{\rm b}(R) = \frac{m_{\rm b}}{4 \pi (R+r_{\rm b})(R^2+r_{\rm b}^2)},
\end{equation}
where $r_{\rm b}$ is the core radius, and the scale mass $m_{\rm b}$ is the mass within the core.  Here we set $r_{\rm b} = p \,r_{\rm s} $, where $p$ is a constant that represents the size of the core as a fraction of the scale radius. Furthermore, we also add a smooth truncation term, resulting in a profile of the form
\begin{equation}\label{eq:modified_burkert}
\rho_{\rm tBurk}(R) = \frac{m_{\rm b}}{4 \pi (R + p \; r_{\rm s})(R^2 + p^2 r_{\rm s}^2)} \left(\frac{r_{\rm t}^2}{R^2 + r_{\rm t}^2}\right),
\end{equation}
where the total mass of the subhalo with this profile is given by
\begin{equation} \label{eq:mb}
m = m_{\rm b} \frac{\tau^2 \left(\pi (p - \tau)^2 + 4 \tau^2 \log\left[\frac{p}{\tau}\right]\right)}{4 (p^4 - \tau^4)}.
\end{equation}
We call this a truncated Burkert (tBurk) profile. Note that for a given $p$, the intrinsic parameters for the tBurk subhalos are the same as for the tNFW ones: $\qq = \{r_{\rm s},r_{\rm t}\}$. This profile is shown in Fig.~\ref{fig:sidm_density_profile}, where we show the tNFW profile and tBurk profile for $p = 0.7$. This choice for $p$ is motivated by the fact that Ref.~\cite{Rocha_2012} finds that for them, $r_{\rm b}$ in Eq.~\eqref{eq:burkert} corresponds to the CDM $r_{\rm s}$ value of $r_{\rm b} = 0.7 r_{\rm s}$. The tBurk profile exhibits a characteristic bump expected in SIDM halos, which is due to the redistribution of mass at the halo center caused by injecting kinetic energy from the outskirts of the halo towards the inner regions \cite{Vogelsberger_2015}. 

Using Eq.~\eqref{eq:modified_burkert} we find an analytic expression for the convergence:
\begin{align}\label{eq:sidm_conv}
\kappa_{\rm tBurk}(x) &= \frac{m_{\rm b}}{8 \pi \Sigma_{\rm crit} r_{\rm s}^2} \; \tau^2 \Bigg\{ \pi \Bigg( \frac{2p \sqrt{\frac{1}{\tau^2 + x^2}}}{p^4 - \tau^4} - \frac{\sqrt{\frac{1}{x^2-p^2}}}{p(\tau^2 + p^2)} \en
& - \frac{\sqrt{\frac{1}{x^2 +p^2}}}{p^3 -p\tau^2} \Bigg) + \frac{2 \arctan \left[\frac{p}{\sqrt{x^2 - p^2}} \right]}{\sqrt{x^2 - p^2}(p^3 + p\tau^2)}- \en
& \frac{2  \tanh^{-1} \left[\frac{p}{\sqrt{p^2 + x^2}} \right]}{\sqrt{x^2 +p^2}(p^3-p \tau^2)} + \frac{4 \tau \tanh^{-1} \left[\frac{\tau}{\sqrt{x^2 +\tau^2}} \right]}{\sqrt{x^2 + \tau^2}(p^4-\tau^4)} \Bigg\},
\end{align}
where again $x= r/r_{\rm s}$ and $\tau = r_{\rm t}/r_{\rm s}$ (refer to Appendix \ref{sect:sidm_conv} for details).

As stated above, we are assuming that the spatial distribution of subhalos within the host dark matter halo remains essentially intact in going from CDM to SIDM. Under this assumption, the 2-subhalo term should remain unchanged in going from one dark matter scenario to the other. Of course, realistically it is likely that the 2-subhalo term would actually be different to some extent: as subhalos orbit the host, the friction felt between the parent halo and the smaller subhalos would have an effect on the correlation of subhalo positions, especially since this effect would affect different subhalo orbits asymmetrically.

Assuming the 2-subhalo term to be the essentially same as in the tNFW case, we focus the rest of this section on the expected redistribution of power on small scales in the 1-subhalo term. In the forthcoming discussion we will therefore explore the extent of this high-$k$ difference between the two density profiles we've chosen to be representative of each dark matter scenario. 

\begin{figure}[t!]
\centering
\includegraphics[width=0.495\textwidth]{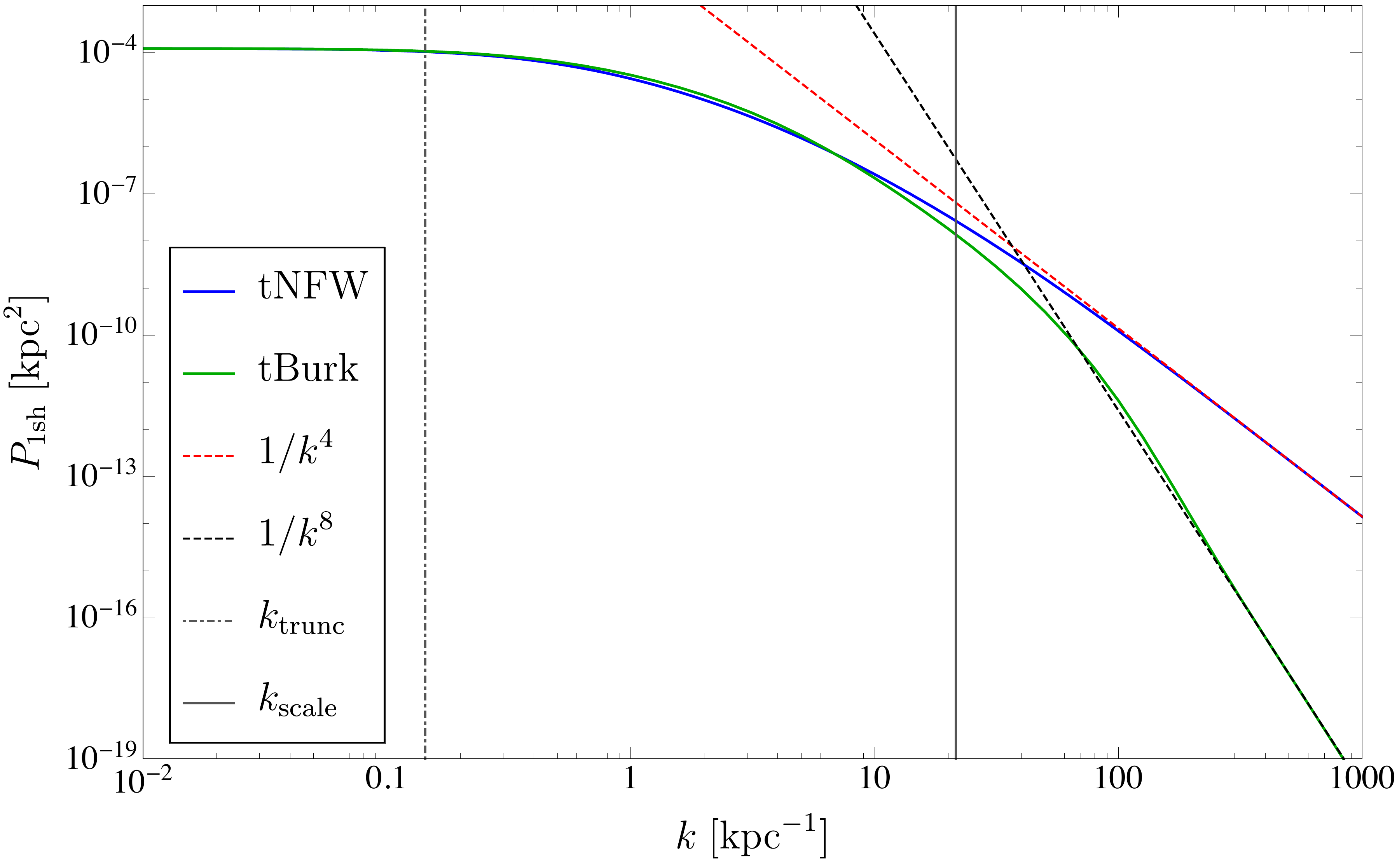}
\caption{1-subhalo power spectrum for a population of tNFW subhalos (solid blue; same fiducial model as in Fig. \ref{fig:powerspec_hezaveh}) and tBurk subhalos (solid green). We also show $k_{\rm trunc}$ (dotted-dashed gray) and $k_{\rm scale}$ (solid gray), as well as the $k\gg k_{\rm scale}$ behavior of both power spectra.}\label{fig:P1sh_SIDM}
\end{figure}
We follow an identical procedure to the tNFW case to determine the 1-subhalo term of the power spectrum, which is shown in Fig.~\ref{fig:P1sh_SIDM}. We also show, for reference, the fiducial tNFW case shown in blue in Fig.~\ref{fig:powerspec_hezaveh}. There is a slight increase in power with respect to the tNFW population on intermediate scales due to the redistribution of mass as the core forms, followed by the expected decrease in power on small scales due to the actual core. Despite these differences, we note that the changes of the substructure convergence power spectrum on scales $k_{\rm trunc}\lesssim k \lesssim k_{\rm scale}$ in going from the tNFW to the tBurk case is well within the variation allowed by varying the statistical properties of the subhalo population, i.e., the different effects shown across Figs.~\ref{fig:powerspec_hezaveh} and \ref{fig:p1sh_ensavg}. This implies that measurements of the power spectrum on these scales are unlikely to distinguish between a cored or cusped subhalo profile.

On even smaller scales $k\gg k_{\rm scale}$, the tBurk power spectrum $P_{\rm 1sh}(k)$ begins to significantly deviate from its tNFW counterpart. Indeed, since the Fourier transform of the truncated Burkert profile behaves as
\be
\tilde{\kappa}_{\rm tBurk}(k) \rightarrow \frac{ 8 (p^4 - \tau^4)}{\tau^2 \left(\pi (p - \tau)^2 + 4 \tau^2 \log\left[\frac{p}{\tau}\right]\right)}\frac{1}{(k\, p\,r_{\rm s})^4},
\ee
for $k\,p\, r_{\rm s}\gg1$, the 1-subhalo term for a population of cored subhalos goes as $P_{\rm 1sh}(k)\propto 1/k^8$ for large $k$, much steeper than the $1/k^4$ expected for NFW subhalos. Therefore, if at all measurable (see discussion below), the slope of the power spectrum on these scales could be decisive in determining the inner density profile of subhalos, which in turn could shed light on the particle nature of dark matter. 

%%%%%%%%%%%%%%%%%%%%%%%%%%%%%%%%%%%%%%%%%%%%%%%%%%%%
\section{Discussion and Conclusion}
\label{sec:conclusions}

In this paper we have introduced a general formalism to study the two-point correlation function of the convergence field due to subhalo populations in strong gravitational lenses, keeping in mind that the observables for these types of problems tend to be photon count or surface brightness maps that exhibit multiple images due to the light from a background source (e.g.~a quasar or a galaxy) having been warped by a massive foreground object, namely the gravitational lens. We have explored in depth how different subhalo population properties affect the substructure convergence field, as well as how it differs for two alternative dark matter scenarios: CDM, which we have represented as a population of tNFW subhalos, and SIDM, where we used a truncated generalized Burkert profile to represent the subhalo population. 

Using the CDM scenario as our baseline, we found that the form of the 1-subhalo term is largely determined by three key quantities: a low-$k$ amplitude proportional to $\bksub \langle m^2\rangle/\langle m \rangle$, a turnover scale $k_{\rm trunc}$ where the power spectrum starts probing the density profile of the largest subhalos, and the wave number $k_{\rm scale}$ corresponding to the smallest scale radii beyond which the slope of the power spectrum reflects the inner density profile of the subhalos. We have shown that the first of these is directly related to subhalo abundance and specific statistical moments of the subhalo mass function. On the other hand, the turnover scale is determined by the average truncation radius of the largest subhalo included in the power spectrum calculation. On scales $k_{\rm trunc} \lesssim k \lesssim k_{\rm scale}$, there is significant variability depending on the statistical properties of subhalos - i.e.~changes to the tidal truncation, to parameters pertaining to the subhalo mass function, or to the scale radius-mass relation can shift the distribution of power and slope on these scales in a rather degenerate manner (see Figs.~\ref{fig:powerspec_hezaveh} and \ref{fig:p1sh_ensavg}). This indicates that measurements of the substructure convergence power spectrum might not be able to distinguish between changes to these different subhalo statistical properties.

For SIDM-like subhalos with a truncated Burkert profile, much of the same discussion applies. While in general the difference between the tNFW and tBurk power spectra is well within the range allowed by varying subhalo population parameters (such as the mass function), there is one defining characteristic that could set both scenarios apart: the high-$k$ slope. For a population of cored, tBurk subhalos, the high-$k$ slope is much steeper than for tNFW, and goes as $1/k^8$ as opposed to the $1/k^4$ behavior of tNFW. While not discussed in this paper, we note that a population of truncated isothermal (``pseudo-Jaffe'', \cite{Keeton2001}) subhalos would lead to a shallower substructure convergence power spectrum going as $1/k^2$ at large wave numbers. Remarkably, the high-$k$ ($k \gtrsim k_{\rm scale}$) slope appears robust to changes in other parameters that govern the statistical properties of the subhalo population, despite the variation at intermediate wave numbers. Note that this is true even when taking into account our lack of knowledge about the 2-subhalo term, since it will not have a noticeable contribution on such small scales. Therefore, determining the high-$k$ slope of the power spectrum would be of particular interest since it would allow us to distinguish between cusped and cored profiles, and more generally, to determine the average small-$r$ behavior of the subhalo density profile. 

The Fisher forecast estimates of Ref.~\cite{Hezaveh_2014} (Fig.~5 in their paper) seem to imply that $\sim 10 - 40$ hour long ALMA observations would be able to measure the amplitude of the power spectrum as well as $k_{\rm trunc}$. However, based on their results, it seems unlikely that these observations would be able to constrain the high-$k$ slope of the power spectrum. Therefore, although we may characterize the abundance of subhalos and the average size of the largest unresolved subhalos with ALMA, it appears unlikely that we will be able to fully constrain the average density profile of subhalos. A measurement of the latter would require a $\sim10$ pc-level resolution within an object that is cosmologically distant from the Milky Way, a very difficult observation indeed, but not necessarily out of reach of very long baseline interferometry. Even if such a measurement could be made, however, it is likely that baryonic structures such as giant molecular clouds \cite{2011ApJ...729..133M} and globular clusters \cite{2012A&ARv..20...50G,He:2017qdo} would contribute to the convergence power spectrum on these scales and could contaminate the signal on scales $k\gtrsim 10$ kpc$^{-1}$. 

There are several potential future directions to the work presented here. An immediate next step would be to compare our analytical results to the substructure convergence power spectrum extracted from high-resolution simulations. Such a comparison could also allow us to obtain a better estimate of the magnitude of the 2-subhalo term, and help us determine whether it can become more important than the 1-subhalo term on larger scales. It would also be interesting to estimate the contribution to the convergence power spectrum from baryonic structures and line-of-sight subhalos \citep{Xu:2009ch,Xu:2011ru,Li:2016afu}. Our analysis could also be improved by allowing the internal shape of the subhalo density profile to vary as a function of mass to take into account the fact that more massive subhalos may be more affected by baryonic feedback (and thus allowing them to form cores) than less massive subhalos. In order to combine measurements from different strong lenses, it will also be of primary importance to understand how the substructure power spectrum depends on the properties (e.g.~redshift, concentration, stellar content, etc.) of the host lens galaxy \cite{Mao:2015yua}.

In this paper, we have computed the lens plane-averaged (that is, the monopole) substructure convergence power spectrum since it is the quantity that is most readily extracted from observations. However, since lens galaxies are generally not spherically symmetric (see e.g.~Refs.~\cite{2013ApJ...777...97S,2017MNRAS.469.1824X}), it is entirely possible that the substructure power spectrum is not isotropic, and it might be fruitful to also consider the higher multipoles of the power spectrum, as it is done, for instance, in the case of the galaxy power spectrum in large-scale structure surveys (see e.g.~Ref.~\cite{Beutler:2016arn}).  By breaking rotational symmetry a new relevant scale could arise in the power spectrum, potentially breaking some of the degeneracy between different astrophysical parameters that was exhibited in the power spectra we considered in this paper. In addition, it is possible that non-Gaussian signatures encoded in the higher $n$-point correlation functions could also contain important information about mass substructures within lens galaxies.  

In conclusion, we have performed a detailed study of the amplitude and shape of the substructure convergence power spectrum within lens galaxies. We have shown how important features of the subhalo population get imprinted on the power spectrum. Based on the sensitivity and resolution of near-future observations, it appears unlikely that substructure power spectrum measurements would be able to probe the inner density profile of dark matter subhalos. Nevertheless, such measurements will provide some constraints on the abundance, mass function, and tidal truncation of low-mass subhalos within lens galaxies, and thus constitute a key consistency test of the standard CDM paradigm. In the event that the measured substructure power spectrum significantly deviates from our CDM expectations, they may even shed new light on the particle nature of dark matter.

%%%%%%%%%%%%%%%%%%%%%%%%%%%%%%%%%%%%%%%%%%%%%%%%%%%%
\acknowledgments
This work was performed in part at the Aspen Center for Physics, which is supported by National Science Foundation Grant No. PHY-1607611. F.-Y. C.-R.~acknowledges the support of the National Aeronautical and Space Administration ATP Grant No. NNX16AI12G at Harvard University.

%%%%%%%%%%%%%%%%%%%%%%%%%%%%%%%%%%%%%%%%%%%%%%%%%%%%
\appendix 

\section{Deriving $\mathcal{P}_{\rm t}$}\label{sect:deriving_pt}

Let us consider a subhalo population that is uniformly distributed. Starting with Eq.~\eqref{eq:integral_pt} and setting $g(h) \equiv r_{\rm t} - r_{\rm t,0} \left(\frac{m}{m_0}\right)^{1/3} \left( \frac{\sqrt{r^2 + h^2}}{r_{\rm 3D,0}} \right)^{\nu}$, we obtain
\begin{align}
\mathcal{P}_{\rm t}(r_{\rm t}|m,r) &= \frac{1}{Z} \int dh \: \mathcal{P}_{3D}(r_{3D}) \; \delta (g(h)) \en
&= \frac{1}{Z} \frac{1}{2 A R_{\rm max}} \int dh \; \delta(g(h)) \en
&=  \frac{1}{Z} \frac{1}{2 A R_{\rm max}} \frac{2}{|g'(h_i)|} \en
&= \frac{1}{R_{\rm max}|g'(h_i)|},
\label{eq:pt_ghi}
\end{align}
where $h_i$ is the solution of $g(h_i)=0$, and where we used $Z = 1/A$. Then,
\begin{align}
|g'(h_i)| &= \nu \frac{r_{\rm t}}{r_{\rm 3D,0}^2} \sqrt{r_{\rm 3D,0}^2 \left[\left(\frac{m_0}{m}\right)^{1/3} \frac{r_{\rm t}}{r_{\rm t,0}} \right]^{2/\nu}- r^2}  \en
& \hspace*{0.75cm} \times \left[\left(\frac{m_0}{m}\right)^{1/3} \frac{r_{\rm t}}{r_{\rm t,0}} \right]^{-2/\nu}.
\end{align}
where $0 \leq r \lesssim b$. Letting $x^2 = r_{\rm 3D,0}^2 \left[\left(\frac{m_0}{m}\right)^{1/3} \frac{r_{\rm t}}{r_{\rm t,0}} \right]^{2/\nu}$, we can do the following expansion:
\begin{align}
\sqrt{x^2 - r^2} = x \sqrt{1 - \frac{r^2}{x^2} + ....} \approx x ,
\end{align}
 where we have used the fact that $x^2 \gg r^2$. In reality this equality does not hold perfectly: when subhalos are at 3D halo-centric distances close to (or below) the Einstein radius, their tidal radius can be such that $x^2$ is comparable to (or less than) $r^2$. However, we can take advantage of the fact that the volume in which $r_{\rm t}$ takes on such small values makes up only $\sim 1\%$ of the entire line-of-sight volume within the host, so the number of subhalos with these tidal radii will make up a minute portion of the entire subhalo population after projection onto the lens plane. 
 
 Then,
\begin{align}
|g'(h_i)| &= \nu \frac{r_{\rm t}}{r_{\rm 3D,0}} \left[\left(\frac{m_0}{m}\right)^{1/3} \frac{r_{\rm t}}{r_{\rm t,0}} \right]^{-1/\nu}
\end{align}
and plugging this into Eq.~\eqref{eq:pt_ghi},
\begin{align}
\mathcal{P}_{\rm t}(r_{\rm t}|m) = \frac{1}{\nu R_{\rm max}} \frac{r_{\rm 3D,0}}{r_{\rm t}} \left[\left(\frac{m_0}{m}\right)^{1/3} \frac{r_{\rm t}}{r_{\rm t,0}} \right]^{1/\nu}.
\end{align}
In fact $\mathcal{P}_{\rm t}$ is unchanged in a case where $\mathcal{P}_{\rm r}$ has some radial dependence. Using as an example $\mathcal{P}_{\rm r}(r) = (1/2 \pi b)(1/r)$, we obtain $\mathcal{P}_{3D}(r)  = (1/4 \pi b R_{\rm max})(1/r)$. Then,
\begin{align}
\mathcal{P}_{\rm t}(r_{\rm t}|m,r) &= \frac{1}{Z} \frac{1}{4 \pi b R_{\rm max}} \int dh \: \frac{1}{r} \; \delta (g(h)) \en
&= \frac{2 \pi b r}{4 \pi b R_{\rm max} r} \frac{2}{|g'(h_i)|} \en
&= \frac{1}{R_{\rm max}|g'(h_i)|}. 
\end{align}

\bigskip

%%%%%%%%%%%%%%%%%%%%%%%%%%%%%%%%%%%%%%%%%%%%%%%%%%%%

\section{SIDM convergence profile}
\label{sect:sidm_conv}

To normalize Eq.~\eqref{eq:modified_burkert} (i.e., determine $m_{\rm b}$), we 
simply integrate the profile out to infinity, which gives us Eq. \eqref{eq:mb}. To obtain the convergence profile, we calculate the projection integral
\be
\kappa_{\rm tBurk}(r) = \frac{1}{\Sigma_{\rm crit}}\int_{-\infty}^{\infty} \rho_{\rm tBurk}(\sqrt{h^2 + r^2}) \; dh,
\ee
where $r$ is the 2D radial coordinate on the lens plane, $h$ is the line-of-sight coordinate, and thus $R = \sqrt{h^2 + r^2}$. We can in fact simplify this expression by doing a slight change of variables. We can rewrite Eq.~\eqref{eq:modified_burkert} as 
\begin{equation}
\rho_{\rm tBurk}(y) = \frac{m_{\rm b}}{4 \pi r_{\rm s}^3} \frac{1}{(p+y)(p^2 + y^2)} \left(\frac{\tau^2}{y^2 + \tau^2}\right),
\end{equation}
where $y = R/r_{\rm s}$ and $\tau = r_{\rm t}/r_{\rm s}$. Then, with $l=h/r_{\rm s}$ and $x = r/r_{\rm s}$,
\begin{align}\label{eq:sidm_kappa}
\kappa_{\rm tBurk}(x) &=  \frac{r_{\rm s}}{\Sigma_{\rm crit}} \int_{-\infty}^{\infty} \rho(\sqrt{l^2 + x^2}) \; dl \en
&= \frac{m_{\rm b}}{2 \pi \Sigma_{\rm crit} r_{\rm s}^2} \; \tau^2 \Bigg\{ \pi \Bigg( \frac{2p \sqrt{\frac{1}{\tau^2 + x^2}}}{p^4 - \tau^4} - \frac{\sqrt{\frac{1}{x^2-p^2}}}{p(\tau^2 + p^2)} \en
& - \frac{\sqrt{\frac{1}{x^2 +p^2}}}{p^3 -p\tau^2} \Bigg) + \frac{2 \arctan \left[\frac{p}{\sqrt{x^2 - p^2}} \right]}{\sqrt{x^2 - p^2}(p^3 + p\tau^2)}- \en
& \frac{2  \tanh^{-1} \left[\frac{p}{\sqrt{p^2 + x^2}} \right]}{\sqrt{x^2 +p^2}(p^3-p \tau^2)} + \frac{4 \tau \tanh^{-1} \left[\frac{\tau}{\sqrt{x^2 +\tau^2}} \right]}{\sqrt{x^2 + \tau^2}(p^4-\tau^4)} \Bigg\}.
\end{align}

%%%%%%%%%%%%%%%%%%%%%%%%%%%%%%%%%%%%%%%%%%%%%%%%%%%%

\begin{widetext}
\section{Table of Constants and Variables}
\label{sect:table}

\begin{table}[htbp]
\centering
\begin{tabular}{  c | c | c }
\hline
\textbf{Constant or Variable} & $\mathbf{Value}$ & \textbf{Description} \\ \hline
$M_{\rm lens}$ & $1.8 \times 10^{12}$ $M_{\odot}$ & Lens mass \\
$R_{\rm max}$ & 409.6 kpc & Maximum radius of the lens \\
$b$ & 6.3 kpc & Einstein radius of the lens \\
$\Sigma_{\rm crit}$ & $3 \times 10^9$ $M_{\odot}$/kpc$^2$ & Critical surface mass density \\
$m_{\rm high}$ & $10^8$ $M_{\odot}$ & Upper bound for the subhalo mass \\ 
$m_{\rm low}$ & $10^5$ $M_{\odot}$ & Lower bound for the subhalo mass \\ 
$m_*$ & $2.52 \times 10^7$ $M_{\odot}$ &  \\
$\beta$ & -1.9 & Subhalo mass function slope \\
$r_{\rm s}$ &  & Subhalo scale radius \\
$r_{\rm s,0}$ & 0.1 kpc &  \\
$r_{\rm t}$ &  & Subhalo tidal radius \\
$r_{\rm t,0}$ & 1 kpc &  \\
$r_{\rm 3D}$ &  & 3D halocentric distance to a subhalo \\
$r_{\rm 3D,0}$ & 100 kpc &  \\
$m_{0}$ & $10^6$ $M_{\odot}$ &  \\
\hline
 \end{tabular}
 \end{table}

\end{widetext}

\bibliography{bib}
\bibliographystyle{apsrev4-1}

\end{document}